\documentclass[12pt,preprint]{aastex}
\usepackage{epsf}

\def\deg{\ifmmode {^\circ}\else {$^\circ$}\fi}
\def\degree{\ifmmode {^\circ}\else {$^\circ$}\fi}
\def\mum{\ifmmode {\rm \mu {\rm m}}\else $\rm \mu {\rm m}$\fi}
\def\arcsec{\ifmmode ^{\prime \prime}\else $^{\prime \prime}$\fi}

\def\inch{\ifmmode ^{\prime \prime}\else $^{\prime \prime}$\fi}
\def\arcmin{\ifmmode ^{\prime}\else $^{\prime}$\fi}

\def\mjup{M$_{\rm J}$}

\def\lstar{$L_\star$}

\def\2470{[24]--[70]}

\def\gyr{g~yr$^{-1}$}

\def\mjup{\ifmmode M_{\rm J} \else $M_{\rm J}$\fi}
\def\qdstar{\ifmmode Q_D^\star\else $Q_D^\star$\fi}
\def\cms{\ifmmode {\rm cm~s^{-1}}\else ${\rm cm~s^{-1}}$\fi}
\def\kms{\ifmmode {\rm km~s^{-1}}\else ${\rm km~s^{-1}}$\fi}
\def\mearth{\ifmmode M_\oplus\else $M_\oplus$\fi}
\def\msun{\ifmmode M_\odot\else $M_\odot$\fi}
\def\mstar{\ifmmode M_\star\else $M_\star$\fi}
\def\gyr{\ifmmode {\rm g~yr^{-1}}\else ${\rm g~yr^{-1}}$\fi}
\def\gcms{\ifmmode {\rm g~cm^{-2}}\else ${\rm g~cm^{-2}}$\fi}
\def\gcmc{\ifmmode {\rm g~cm^{-3}}\else ${\rm g~cm^{-3}}$\fi}
\def\rearth{\ifmmode R_\oplus\else $R_\oplus$\fi}
\def\rhill{\ifmmode R_H\else $R_H$\fi}
\def\rhille{\ifmmode R_H^{\prime}\else $R_H^{\prime}$\fi}
\def\mesc{\ifmmode M_{esc}\else $M_{esc}$\fi}
\def\resc{\ifmmode R_{esc}\else $R_{esc}$\fi}
\def\mtot{\ifmmode M_{tot}\else $M_{tot}$\fi}
\def\mlr{\ifmmode M_{LR}\else $M_{LR}$\fi}
\def\rlr{\ifmmode R_{LR}\else $R_{LR}$\fi}
\def\flr{\ifmmode f_{LR}\else $f_{LR}$\fi}
\def\mlf{\ifmmode M_{LF}\else $M_{LF}$\fi}
\def\rlf{\ifmmode R_{LF}\else $R_{LF}$\fi}
\def\flf{\ifmmode f_{LF}\else $f_{LF}$\fi}
\def\rmin{\ifmmode R_{min}\else $R_{min}$\fi}
\def\rmax{\ifmmode R_{max}\else $R_{max}$\fi}
\def\rmaxc{\ifmmode R_{max,c}\else $R_{max,c}$\fi}
\def\rmaxd{\ifmmode R_{max,d}\else $R_{max,d}$\fi}

\def\nd{\ifmmode N_{d}\else $N_{d}$\fi}
\def\md{\ifmmode M_{d}\else $M_{d}$\fi}
\def\mdmin{\ifmmode M_{d,min}\else $M_{d,min}$\fi}
\def\mdmax{\ifmmode M_{d,max}\else $M_{d,min}$\fi}
\def\ab{\ifmmode A_{b}\else $A_{b}$\fi}
\def\ad{\ifmmode A_{d}\else $A_{d}$\fi}
\def\ado{\ifmmode A_{d,o}\else $A_{d,o}$\fi}
\def\adt{\ifmmode A_{d,t}\else $A_{d,t}$\fi}

\begin{document}

\title{Fomalhaut b as a Cloud of Dust: Testing Aspects of Planet Formation Theory} 
\vskip 7ex
\author{Scott J. Kenyon}
\affil{Smithsonian Astrophysical Observatory,
60 Garden Street, Cambridge, MA 02138} 
\email{e-mail: skenyon@cfa.harvard.edu}

\author{Thayne Currie}
\affil{Department of Astronomy \& Astrophysics, University of Toronto, 
50 St. George Street, Toronto, ON M5S 1A1, Canada}
\email{email: currie@astro.utoronto.ca}

\author{Benjamin C. Bromley}
\affil{Department of Physics, University of Utah, 
201 JFB, Salt Lake City, UT 84112} 
\email{e-mail: bromley@physics.utah.edu}
%
%

\begin{abstract}

We consider the ability of three models -- impacts, captures, and collisional cascades -- 
to account for a bright cloud of dust in Fomalhaut b. Our analysis is based on a novel 
approach to the power-law size distribution of solid particles central to each model. 
When impacts produce debris with (i) little material in the largest remnant and (ii) a 
steep size distribution, the debris has enough cross-sectional area to match observations 
of Fomalhaut b. However, published numerical experiments of impacts between 100~km objects
suggest this outcome is unlikely.  If collisional processes maintain a steep size 
distribution over a broad range of particle sizes (300~\mum\ to 10~km), Earth-mass planets 
can capture enough material over 1--100~Myr to produce a detectable cloud of dust. 
Otherwise, capture fails.  When young planets are surrounded by massive clouds or disks 
of satellites, a collisional cascade is the simplest mechanism for dust production in 
Fomalhaut b.  Several tests using HST or JWST data -- including measuring the 
expansion/elongation of Fomalhaut b, looking for trails of small particles along 
Fomalhaut b's orbit, and obtaining low resolution spectroscopy -- can discriminate 
among these models.

\end{abstract}


\section{INTRODUCTION}
\label{sec: intro}

Fomalhaut b is a faint object orbiting at a distance of $\sim$ 120~AU from 
the nearby A-type star Fomalhaut.  Originally detected on {\it Hubble Space
Telescope} (HST) images at 0.6~\mum\ and 0.8~\mum\ \citep{kalas2008}, the 
source lies inside the orbits 
of a bright belt of dust particles at 130--150~AU from the central star 
\citep[e.g.,][]{holl2003,stap2004,kalas2005,marsh2005,ricci2012,acke2012,boley2012,su2013}. 
Recent re-analyses of the original HST data confirm the detections at 
0.6--0.8~\mum\ and identify the source on images at 
0.435~\mum\ \citep{currie2012,galich2013}.  New optical HST data 
recover the object in 2010--2012 \citep{kalas2013}.  In all of these 
studies, the optical colors are similar to those of the central star.

Despite the robust optical data, Fomalhaut b is not detected at infrared 
(IR) wavelengths. Attempts to identify the source have failed at 
1.25~\mum\ \citep{currie2012}, 1.6~\mum\ \citep{kalas2008,currie2013}, 
3.6--3.8~\mum\ \citep{kalas2008,maren2009}, and 4.5~\mum\ \citep{maren2009,janson2012}.  
Each upper limit lies well above the IR fluxes expected for an object 
with the optical-infrared colors of an A-type star. Although the brighter 
IR fluxes expected from a 2--10~\mjup\ (Jupiter mass) planet are excluded 
by these data, the IR data are consistent with emission from lower mass 
planets \citep[e.g.,][]{janson2012,currie2012}.  However, the measured 
optical fluxes in Fomalhaut b are a factor $\gtrsim$ 100 larger than expected 
for a 1~\mjup\ planet at a distance of 7.7 pc from the Earth 
\citep[e.g.,][]{currie2012}. Thus, the optical flux requires a different source.

Without a clear IR detection, the simplest explanation for the emission 
from Fomalhaut b is scattered light from an ensemble of dust grains with 
a collective cross-sectional area of roughly $10^{23}$ cm$^2$ 
\citep[e.g.,][]{kalas2008}.  A single, high velocity collision between 
two objects with radii of 10--1000~km is a plausible source for the dust
\citep{wyatt2002,kb2005,kalas2008,galich2013,kalas2013}. In this picture, 
the collision disperses objects with sizes ranging from a fraction of a micron 
to tens of meters or kilometers 
\citep[see, for example, the discussions in][]{wyatt2002,kb2005}. 

A collisional cascade within a circumplanetary cloud \citep{kw2011a} 
or debris disk \citep{kalas2008} is another plausible source of dust grains 
in Fomalhaut b \citep[e.g.,][] {currie2012,galich2013,kalas2013}.  
In this model, dynamical processes place a massive cloud of satellites
around a newly-formed 1--100~\mearth\ planet \citep[e.g.,][]{nesv2007}. 
Subsequent collisions among 1--100~km satellites produce copious amounts of dust 
\citep[e.g.,][]{bottke2010,kw2011a}. Aside from the nature of the collisions, 
this mechanism probably produces dust grains with properties similar to those 
derived from a single giant impact.

Material continuously captured from Fomalhaut's circumstellar disk provides 
a third source for dust in Fomalhaut b. In this picture 
\citep[e.g.,][]{ruskol1961,ruskol1963,ruskol1972}, material orbiting 
Fomalhaut loses energy and is captured by a massive planet. Collisions
among captured objects produce a cloud of dust grains orbiting the
planet.  If the grains within this disk remain small, their properties 
are probably similar to dust produced in a single collision or in a
collisional cascade.

In this paper, we develop a framework for analyzing dusty clouds of debris 
and apply this framework to available data for Fomalhaut b. We begin in \S2
with a summary of relevant data for this system. In \S3, we consider a novel
approach for deriving properties of the debris from the observed cross-sectional 
area (\S3.1) and apply this approach to dust produced in a giant impact (\S3.2),
captured from the protoplanetary disk (\S3.3), and generated in a collisional
cascade (\S3.4). After exploring uncertainties, tests, and improvements of these 
mechanisms for dust production (\S4), we conclude with a brief summary (\S5).

\section{RELEVANT OBSERVATIONS}
\label{sec: data}

To develop robust models for dust emission in Fomalhaut b, we first establish 
pertinent observational results from existing data. Fomalhaut is a 200--400~Myr old 
A3~V star at a distance, $D$ = 7.7 pc \citep[e.g.,][]{barrado1998,mama2012}.  The 
star has two nearby, apparently bound companions, TW PsA (K4~V) and LP~876-10 (M4~V), 
at distances 0.28--0.77 pc from the primary star \citep{barrado1998,mama2012,mama2013}.
Fomalhaut and LP~876-10 have bright debris disks \citep{gill1986,kenn2013}.  
Fomalhaut b has an eccentric orbit (e $\approx$ 0.8) around Fomalhaut with a semimajor
axis, $a_b \approx$ 160--180~AU \citep{kalas2013,beust2014}. This orbit might intersect 
the orbits of material in the outer debris belt of Fomalhaut \citep{kalas2013,beust2014}.

\subsection{Fomalhaut Debris Disk}

All three dust models depend on the amount of circumstellar material along Fomalhaut b's
orbit. The main belt at 130--150~AU lies outside the current position of Fomalhaut b 
\citep{kalas2005,kalas2013}. Within the belt, the mass in solids is at least
20--40~\mearth\ \citep[e.g.,][]{wyatt2002,holl2003} and is probably less than 
300~\mearth\ \citep{kalas2013}.  Models which fit images and the spectral energy 
distribution suggest that the belt of dust at 130--155 AU contains roughly twice 
as much dust as the region from 35--130~AU \citep[e.g.,][]{acke2012}.  Accounting 
for the difference in surface area, the surface density of dust in the inner disk 
is roughly 15\% of the surface density in the main belt.  

To place these results in the context of planet formation theory, the surface density of 
a protoplanetary disk around Fomalhaut is conveniently parameterized as \citep[e.g.,][]{youdin2013}:
\begin{equation}
\Sigma = d ~ \Sigma_0 \left ( {a \over a_0} \right )^{-p} ~ ,
\label{eq: sigma}
\end{equation}
where $\Sigma_0$ is the initial surface density of solid material at $a = a_0$, 
$p \approx$ 1--2, and $d$ = 0--1 is a depletion factor which accounts for the loss of 
material throughout the evolution of the disk \citep[e.g.,][]{will2011,and2013}.  We 
adopt $\Sigma_0$ = 30~g~cm$^{-2}$, $a_0$ = 1~AU, and $p$ = 1 \citep[e.g.,][]{will2011}. 
These parameters imply an initial mass of 150~\mearth\ at 130--150~AU for $d$ = 1, 
which is reasonably consistent with observations. Thus, we adopt $d$ = 1 for the belt.

Observations suggest a significant depletion of material inside the belt. We assume
that the ratio of dust mass at 35--130 AU to the belt mass at 130--150 AU corresponds
to the current mass ratio for all solids in the disk. With $\Sigma \propto a^{-1}$, 
the initial disk mass from 35--130~AU is roughly five times the mass from 130--150~AU.
If the belt now contains roughly twice the mass as the 35--130~AU region, then the
depletion factor at 35--130~AU is roughly $d \approx$ 0.1.

\subsection{Spectral Energy Distribution of Fomalhaut \lowercase{b}}

\begin{figure}
\centering
\includegraphics[width=6.5in]{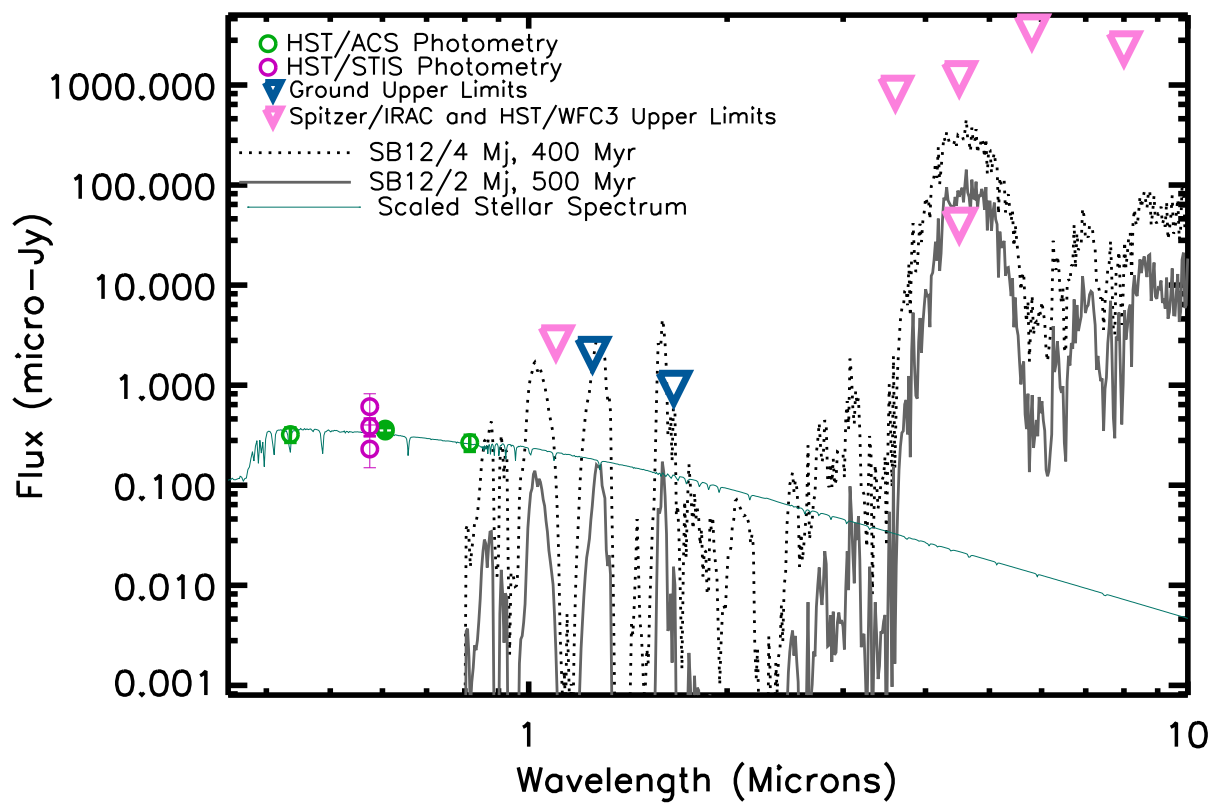}
\caption{SED of Fomalhaut b (open circles and triangles) compared to a scaled-down 
version of the stellar spectrum and synthetic planet spectra (2--4 $M_{J}$, age = 
400--500 $Myr$) from \citet{Spiegel2012}.  In addition to the data listed in Table~1, 
we plot the $STIS$ photometry from \citet{kalas2013} (lower magenta circle) and 
\citet{galich2013} (upper magenta circle).
}
\label{sed}
\end{figure}

To assign a cross-sectional area and size to the dust emission in Fomalhaut b, we
collect existing data for the spectral energy distribution (SED) and the spatial 
extent of the image.  For the SED, we adopt published results from HST detections 
\citep{kalas2008,currie2012,galich2013,kalas2013} and from 
IR upper limits \citep{maren2009,janson2012,currie2012,currie2013}.  Figure \ref{sed} 
compares these data (Table~1) to predictions from model planet atmospheres and a 
scaled-down version of the stellar spectrum.  In addition to data from Table~1, we 
include entries for $STIS$ photometry from \citet{kalas2013} (bottom magenta circle), 
\citet{galich2013} (top magenta circle), and \citet{currie2012} (middle magenta circle). 

The IR non-detections at 1--5~\mum\ place strong constraints on thermal emission from
a Jupiter-mass planet around a 200--400~Myr A-type star.  The near-IR upper limits 
rule out planets more massive than $\sim$ 4--5 \mjup\ \citep[e.g.,][]{currie2013}.  The 
$IRAC$ 4.5~\mum\ data allow masses less than 2 \mjup\ \citep[e.g.,][]{janson2012,currie2012}.

The optical flux density measurements track a scaled down version of the Fomalhaut 
stellar photosphere \citep{currie2012}.  The three HST $ACS$ points in green closely 
follow the photosphere. At 0.6~\mum, three independent reductions of $STIS$ photometry bracket 
the $ACS$ point.  Given the errors, the three $STIS$ measurements agree reasonably well. 

Together, the optical and IR data for Fomalhaut b strongly favor scattered light from 
dust over thermal emission from a Jupiter mass planet. In the optical, the flux is 
more than a factor of $100$ brighter than the emission expected from a planet and has 
the colors of an A-type star. In the IR, the upper limits on the flux density rule out 
planets more massive than 2 \mjup\ and lie a factor of ten brighter than the emission
expected from scattered light.

\subsection{Spatial Extent of Fomalhaut b}

Placing limits on the emitting area of Fomalhaut b requires an understanding of the
HST point-spread-function (PSF) and the noise in $ACS$ and $STIS$. Several published
results suggest the source is unresolved \citep{kalas2008,currie2012}. Others 
report the source is extended. \citet{galich2013} suggest the source is resolved in 
the F814W data; \citet{kalas2013} attribute extended structure in the $STIS$ data to
speckle noise.

To illustrate the difficulty in measuring the spatial scale of the dust in Fomalhaut b, 
we re-derive the PSF along the $x$ and $y$ axes of the F435W, F606W, and F814W reductions 
of $ACS$ data from \citet{currie2012}.  We try two approaches. First, we construct radial 
intensity profiles in the $x$ and $y$ directions, re-sample the profile with a grid spacing of 0.25 
pixels using linear interpolation, and measure the full-width-at-half-maximum (FWHM) using
a minimum uncertainty of 1/2 a pixel ($\sim$ 13 mas).  Second, we model the intensity 
profile as a 2D gaussian, using the \textit{mpfit} package and adopt the average of results 
for the FWHM from a range of fit radii. Here, we include the standard deviation of these 
measurements in our uncertainty.  For the highest-quality data (F606W), we derive the FWHM 
from both the the 2004 data and the 2006 data, averaging the results for two separate 
reductions of each data set.

Table~\ref{fwhmest} lists our results along with predictions for an unresolved point source.  
In the highest-quality data sets (2004 and 2006 F606W), Fomalhaut b is clearly consistent 
with a point source.  At F435W, Fomalhaut b is slightly extended along the $y$ axis compared 
to a point source.  However, this deviation is barely larger than 1-$\sigma$ and thus is not 
significant. The azimuthally-averaged FWHM (63 $\pm$ 20 mas; 65 $\pm$ 22 mas) is consistent 
with the point source. 

\begin{figure}
\centering
\includegraphics[width=6.5in]{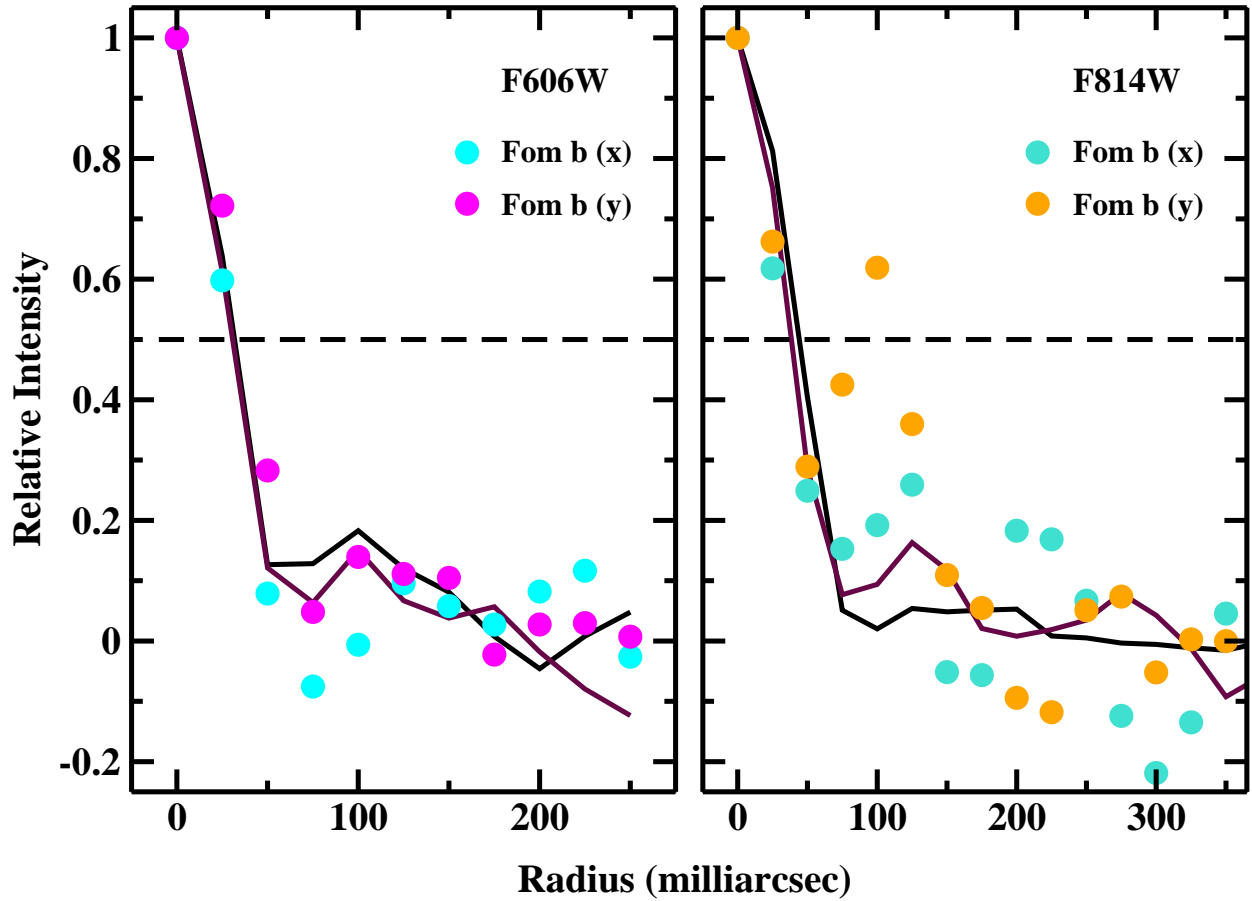}
\vskip 3ex
\caption{Normalized radial intensity profiles along the $x$ and $y$ axes 
for Fomalhaut b for F606W (left panel) and F814W (right panel). In each
panel, black (maroon) lines indicate the intensity profile for a background 
star along the $x$ ($y$) axis.  Fomalhaut b's radial intensity 
profile at F606W is consistent with the point source.  
The profile at F814W is more difficult to interpret.
}
\label{fwhmplot}
\end{figure}

Figure \ref{fwhmplot} demonstrates the necessity of higher SNR data with well-sampled PSFs 
to assess the spatial extent of Fomalhaut b. In both panels, a background star (black and 
maroon lines) has a sharp core with FWHM $\approx$ 35 milliarcsec and a faint halo extending 
to roughly 200 milliarcsec. Within the errors, the $x$ and $y$ traces are indistinguishable.
In F606W (left panel), the $x$ and $y$ traces of Fomalhaut b closely follow results for the 
point source.  Although the F814W data (right panel) have a similarly sharp core inside 50 
milliarcsec, both traces have much larger intensity than a point source at 100--200 milliarcsec. 
Based on this larger intensity, \citet{galich2013} conclude the source is extended. However, 
both traces also have several maxima, suggesting a significant noise component. 

We conclude that Fomalhaut b is unresolved at F435W and at F606W. At F814W, current results are
inconclusive due to the lower SNR relative to F606W. Adopting an angular diameter of 69 $\pm$ 14 
milliarcsec from the highest SNR data (2006 F606W), an upper limit on Fomalhaut b's projected 
radius is $R_b \lesssim R_{b,max} \approx$ 0.5~$\theta ~ D$ $\approx$ 0.27 $\pm$ 0.05 ($D$ / 7.7 pc) AU.  

\subsection{Limits on Emitting Area and Mass}

Following the approach of \citet{kalas2008}, we estimate \ab\ the cross-sectional 
area of dust in Fomalhaut b.  The flux received from the star at the Earth is 
$f_{\star} = L_{\star} / 4\pi D^{2}$.  Fomalhaut b intercepts 
a fraction of the stellar flux $f_b = L_{\star}/4 \pi r^{2}$, where $r$ = 120~AU. 
The observed flux from Fomalhaut b at Earth, $f_o = f_b \ab\ Q_s / 4 \pi D^{2}$, 
depends on the cross-sectional area \ab\ and the scattering efficiency $Q_{s}$. 
Thus, $f_o$ = $(f_\star / 4 \pi r^2) \ab\ Q_s$; 
$\ab = (4 \pi r^2 / Q_s) (f_o / f_{\star})$.  

Deriving \ab\ requires three measured quantities, $r, f_o, f_\star$, and one adopted 
quantity, $Q_s$. For the ratio $f_o / f_\star$, we define the contrast in optical
magnitudes $\Delta m$ = -2.5~log~($f_o$/$f_{\star}$). Adopting $m$ = 1.2 for the primary 
and $m$ = 24.95 for Fomalhaut b, the cross-sectional area for $r \approx$ 120~AU is 
\begin{equation}
\ab = 1.3 \times 10^{23} \left ( { 0.1 \over Q_{s}} \right ) ~ {\rm cm^{2}} ~ .
\end{equation}
This expression assumes grains with albedo similar to objects in the outer solar system
\citep[$Q_{s} \approx$ 0.1;][]{stansberry2008}.  Our estimate is midway between 
previous results of $\ab \approx 10^{23} (0.1 / Q_s)$~cm$^2$ \citep{kalas2008} 
and $\ab \approx 1.5 \times 10^{23} (0.1 / Q_s)$~cm$^2$ \citep{galich2013}. 

For simplicity, we adopt $\ab \approx 10^{23}$~cm$^2$.  A spherical object with
this \ab\ has a radius, $R_b \gtrsim R_{b,min} \approx$ 10$^{11}$ cm $\approx$ 
150 \rearth, somewhat larger than a solar radius and significantly larger than 
the radius of any planet.

If a dust cloud gravitationally bound to a planet produces the observed emission 
in Fomalhaut b, the size and emitting area constrain the mass of the planet 
\citep[e.g.,][]{kalas2008,kw2011a,galich2013,kalas2013}.  For planets orbiting 
a star, material inside the Hill sphere is bound to the planet. The radius of 
the Hill sphere for a circular orbit is
\begin{equation}
\rhill = a \left ( { M_p \over 3 \mstar } \right)^{1/3} ~ ,
\label{eq: rhill}
\end{equation}
where $a$ is the semimajor axis of the planet.  When planets have eccentric orbits, 
$\rhille \approx (r / a) \rhill$, where $r$ is the current distance from the planet
to the star.  With $r$ varying between $a (1 - e)$ at periastron to $a (1+e)$ at 
apoastron, \rhille\ at periastron is $1-e$ smaller than \rhill\ \citep{ham1992}. 
Setting $R_b$ equal to \rhille, a rough limit on the mass of the planet is
\begin{equation}
M_p \approx 2 \left ( { R_b \over 0.01 ~ r } \right )^3 ~ \mearth ~ .
\label{eq: mp}
\end{equation}

The mass of the planet is very sensitive to $R_b$ \citep[][and references therein]{kw2011a,kalas2013}.  
For a barely resolved Fomalhaut b with radius $R_b \approx R_{b,max} \approx $~0.25~AU at 
$r \approx$ 35~AU, $M_p \approx$ 0.8~\mearth.  If Fomalhaut b is optically thick, the 
minimum physical size of the cloud is roughly 0.01~AU. A strong lower limit on the mass
of the planet is then roughly $10^{23}$~g \citep{kalas2008,kw2011a}.  Adopting
$R_b \approx \gamma \rhille$ with 
$\gamma \approx$ 0.2--0.3 \citep[where orbits are definitely stable, e.g.,][]{ham1992} to
$\gamma \approx$ 2--3 \citep[where some orbits are stable for long periods, e.g.,][]{shen2008} leads 
to a much broader range of plausible planet masses \citep[e.g.,][]{kalas2008,kw2011a,kalas2013}.

\subsection{Limits on Optical Depth}

Previous studies of dust emission in Fomalhaut b focus on optically thin models
\citep[e.g.,][]{kalas2008,kw2011a,galich2013,kalas2013}.  A robust lower limit on 
the optical depth $\tau$ depends on the emitting area and the spatial extent. With
\ab\ $\approx$ $10^{23}$~cm$^2$ and $R_b \lesssim$ 0.25~AU, 
$\tau \approx \ab / \pi R_b^2$ $\gtrsim 2 \times 10^{-3}$. 

To derive an upper limit on $\tau$, we assume a cloud composed of particles with mass 
density $\rho$ and radius $R$. The swarm has radius $R_b \approx 10^{11}$~cm and total
mass $M_b \approx \ab \rho R$.  If the cloud is produced in a giant impact, the 8~yr 
baseline of the HST observations establishes a maximum expansion velocity of roughly 300~\cms. 
Setting this velocity equal to the escape velocity of a colliding pair of icy planetesimals 
with mass density $\rho_p \approx$ 2~\gcmc\ yields a planetesimal radius $R_p \approx$ 5~km 
and mass $M_p \approx 5 \times 10^{17}$~g. Requiring $M_p \approx M_b$ yields a typical
particle size, $R \approx M_p / \ab \approx$ 0.05~\mum. Even in an optically thick cloud,
radiation pressure rapidly accelerates such small particles to velocities much larger
than 300~\cms\ \citep[e.g.,][]{burns1979}. Thus, an optically thick cloud from a giant 
impact cannot produce the observed scattered light emission from Fomalhaut b.

We now examine the possibility of an optically thick particle cloud orbiting a 
massive planet.  Particles collide with a collision time $t_c$. In every collision, there 
is some dissipation of the collision energy.  Over time, repeated dissipative collisions 
produce a flattened structure with a finite scale height set by the particle size and 
the semimajor axis of an orbit \citep[e.g.,][]{brah1976}. Collisions eject some particles 
from the system; others fall onto the planet. The optical depth declines.

The collision time for this process is $t_c \approx (n \sigma v)^{-1}$, where $n$ 
is the number density of particles, $\sigma$ is the cross-section, and $v$ is the 
relative velocity. For planets with mass $M_p \approx$ 0.1--10~\mearth, radiation 
pressure sets a minimum particle size $R \approx$ 100~\mum\ \citep[e.g.,][]{kw2011a}.
The number density is $n \gtrsim 10^{-6}$~cm$^{-3} (R / 100~\mum)^{-2}$.  Setting 
$v$ equal to the orbital velocity around the planet \citep[e.g.,][]{kw2011a}, the 
collision time depends only on the mass of the central planet:
\begin{equation}
\label{eq: tcoll-thick}
t_c \lesssim 0.01 \left ( { 1~\mearth\ \over M_p } \right)^{1/2} ~ {\rm yr} ~ .
\end{equation}
For any plausible planet mass, the collision time is 9--11 orders of magnitude smaller 
than the age of Fomalhaut. Thus, the optical depth of an optically thick cloud declines 
on time scales much shorter than the age of Fomalhaut.  

Collision outcomes cannot change this conclusion. If collisions produce larger 
merged objects, the optical depth declines more rapidly. If collisions produce
clouds of smaller particles, radiation pressure removes these particles on 
(i) the time scale for particles to orbit the planet, $t \approx$ 1~yr 
($R \lesssim$ 5--10~\mum) or
(ii) the time scale for the planet to orbit the central star, $t \approx$ 1000~yr
($R \approx$ 10--100~\mum). Both of these time scales are much shorter than the 
age of Fomalhaut. 

This analysis suggests that Fomalhaut b is not a massive, optically thick cloud 
of small particles. For a cloud expanding from a giant impact, the particle
size (0.05~\mum) required for the derived mass is too small. If the cloud orbits
a massive planet, the collision time is too short. Thus, we focus on optically
thin models for the dust emission.

\section{DUST MODELS}

In giant impact models, two large protoplanets collide to produce an ensemble 
of objects with a broad range of sizes \citep[e.g.,][]{wyatt2002,kb2005}. After 
the collision, the center-of-mass of the ensemble -- which might contain a 
few massive objects -- follows an orbit with angular momentum per unit mass 
comparable to the sum of the angular momenta of the two protoplanets. Other 
objects expand away from this orbit; smaller particles expand faster than 
larger particles.  Although the initial expansion of the cloud is roughly 
spherical, orbital shear and collisions with other particles change the shape 
and the mass of the cloud on orbital time scales.  After 10--20 orbital periods 
($\sim 10^4$~yr for Fomalhaut b), the material lies in a narrow ring surrounding 
the central star. 

Collisional cascades begin with a massive, roughly spherical \citep[e.g.,][]{bottke2010,kw2011a} 
or disk-shaped \citep{kalas2008} swarm of satellites orbiting a massive planet.
Destructive collisions among the satellites produce copious amounts of debris 
\citep{bottke2010,kw2011a}.  Collisions within the debris yield even smaller 
particles. The resulting cascade of collisions slowly grinds small satellites 
into dust \citep[e.g.,][]{dohn1969,will1994,tanaka1996b,obrien2003,koba2010a}.  
Radiation pressure and Poynting-Robertson drag remove small dust particles from 
circumplanetary orbits \citep{burns1979}. Thus, the collisional cascade gradually 
removes material from the system. The time scale for the cascade is usually 
10--100~Myr, much longer than the lifetime of material produced in a single, 
giant impact.

Continuous capture models combine aspects of both approaches 
\citep[e.g.,][]{ruskol1972,weiden2002,estrada2006,koch2011}. 
In this picture, a massive planet lies embedded within a circumstellar disk. 
When circumstellar objects pass through the Hill sphere of the planet, they can
lose energy through dynamical interactions with other objects outside the Hill
sphere or through collisions with other objects inside the Hill sphere. If the 
energy loss is large enough, these objects become bound to the planet. Over time,
high velocity collisions between the captured objects lead to the production of
small dust grains. If collisions are fairly frequent and the net angular momentum 
of captured objects is large enough, collisional damping leads to the formation 
of a circumplanetary disk \citep{brah1976}. Otherwise, captured satellites lie 
in a roughly spherical cloud around the planet. 

The evolution of solids within a captured cloud or disk depends on the accumulation 
rate.  Here, we distinguish between the relatively rapid capture of a massive swarm
of satellites during the early evolution of the planetary system \citep[e.g.,][]{nesv2007}
from the slow capture of material throughout the evolution of the planetary system
\citep[e.g.,][]{ruskol1972,weiden2002,estrada2006,koch2011}.
Prompt captures over a few Myr enable the immediate onset of a collisional cascade and
formation of a massive dust cloud. Over time, this evolution may produce an irregular 
satellite system similar to those surrounding the giant planets of the solar system
\citep[e.g.,][]{bottke2010,kw2011a}. When captures occur intermittently, the mass in
satellites grows slowly with time. As this mass grows, collisions gradually produce 
a cloud of debris. Thus, the time scale to produce an observable dust cloud is much 
longer.  In less massive systems composed of small particles, some circumstances allow 
the particles to avoid collisions \citep[e.g.,][]{heng2010}. For any outcome, the 
lifetime of the cloud or disk is 100~Myr or longer.

To isolate important issues in capture and cascade models, we examine two extreme
cases. For cascades, we follow \citet{kw2011a} and assume an initially massive cloud 
of satellites where destructive collisions and radiation pressure slowly reduce the 
mass with time.  The ability of a cascade to match observations of Fomalhaut b then 
depends on the mass of the planet, the initial mass and size of the cloud, and the 
typical particle size \citep[\S3.4; see also][]{kw2011a}.  For captures, we assume 
the initial mass of the cloud is zero and derive the capture rate for objects passing 
through the Hill sphere. Because the capture rate depends on the properties of the
circumstellar disk and the planet (\S3.3), the conditions required for successful
capture models differ from those of cascade models. By focusing on the two models 
separately, we can place better limits on the source of material involved in either 
mechanism.

Aside from the lifetime, various observations might distinguish between these 
dust formation processes. Developing these constraints requires clear predictions
for the mass, cross-sectional area, and other properties of the debris as a function 
of initial conditions and time.  In the next sections, we derive basic properties 
of the debris expected from each model and compare our results with observations 
of Fomalhaut b. Our goal is to develop a better analytic understanding of each 
mechanism which will serve as the foundation for detailed numerical simulations
in future studies. 

\subsection{Properties of the Debris}

To establish the basic properties of a dusty cloud or disk of debris for 
Fomalhaut b, we consider an ensemble of solid particles with total mass \md\ and 
total cross-sectional area\footnote{Throughout the text, we use cross-sectional
area and area interchangeably and reserve {\it surface area} for the total surface 
area of the swarm of particles within the cloud. The total surface area is four 
times larger than the cross-sectional area.} \ad.  In most applications, the
smallest particles have most of the area; the largest particles have most of the 
mass. Matching observations then requires (i) setting an appropriate size for the 
smallest particles, (ii) adopting a size distribution, and (iii) verifying that 
the largest particles contain a reasonable amount of mass. In this paper, our goal 
is to predict the range of particle sizes for specific theories of dust production 
and to learn whether these predictions match observations.  With improved constraints, 
we develop a better understanding of the applicability and limitations of each theory.

To relate the area to the physical radii $R$ of the particles, we assume a size 
distribution $n(R)$, where the number of particles with radii between $R$ and 
$R+dR$ is a power law:
\begin{equation}
n(R) dR = n_0 R^{-q} dR ~ . 
\label{eq: size-dist}
\end{equation}
The total number of particles between a minimum size \rmin\ and a maximum size 
\rmax\ is \nd. 

Here, we require that the number of particles with $R \ge \rmax$ is exactly 1.  
Integrating the size distribution from \rmax\ to infinity and adopting $q > 1$:
\begin{equation}
n_0 = (q - 1) \rmax^{q-1} ~ .
\label{eq: n0}
\end{equation}
For typical $q \approx$ 3.5--6 \citep[e.g.,][]{dohn1969,obrien2003,koba2010a,lein2012}, 
it is very likely that the particle with $R \ge \rmax$ has a radius \rmax. Thus, 
we can integrate over the size distribution from \rmin\ to \rmax\ to derive the 
cross-sectional area:
\begin{equation}
\ad = \pi (q - 1) \rmax^2 \left\{
\begin{array}{lll}
{\rm ln}(\rmax / \rmin) & & q = 3 \\
\\
((\rmax / \rmin)^{q-3} - 1) / (q - 3) ~ & & q \neq 3 \\
\end{array}
\right.
\label{eq: area}
\end{equation}
For all $q > 3$, the smallest particles contain most of the area. 
The total mass requires a similar integral:
\begin{equation}
\md = {4 \pi \rho \over 3} (q - 1) \rmax^3 \left\{
\begin{array}{lll}
{\rm ln}(\rmax / \rmin) & & q = 4 \\
\\
((\rmax / \rmin)^{q-4} - 1) / (q - 4) ~ & & q \neq 4 \\
\end{array}
\right.
\label{eq: mass}
\end{equation}
where $\rho$ is the mass density of the particles.

Our goal is to predict \ad\ and \md\ for each model and to identify combinations 
of model parameters where the predictions match the observed \ab.
Formally, we should augment \ad\ and \md\ by the cross-sectional area and the
mass of the single object with $R \ge \rmax$. With a measured \ad\ $\approx 10^{23}$~cm$^2$,
the correction is less than one part in $10^6$ and safely ignored. For $q > 4$,
most of the mass is in the smallest objects; thus, the largest object makes a
negligible contribution to \md. For small $q \approx$ 3.5, a single object with
$R = \rmax$ adds roughly 1\% to the mass. This correction is negligible.

In these expressions, \rmax\ sets the basic level for the mass and the area. 
The terms involving $\rmax / \rmin$ to the right of the left curly bracket then 
provide a scale factor.  For $q \lesssim$ 4 and any $\rmax/\rmin$, the scale 
factor for the mass is negligible. For $q \gtrsim 4$ ($q \gtrsim$ 3), the scale 
factor for the mass (cross-sectional area) is very sensitive to $\rmax/\rmin$.

\begin{figure}
\includegraphics[width=6.5in]{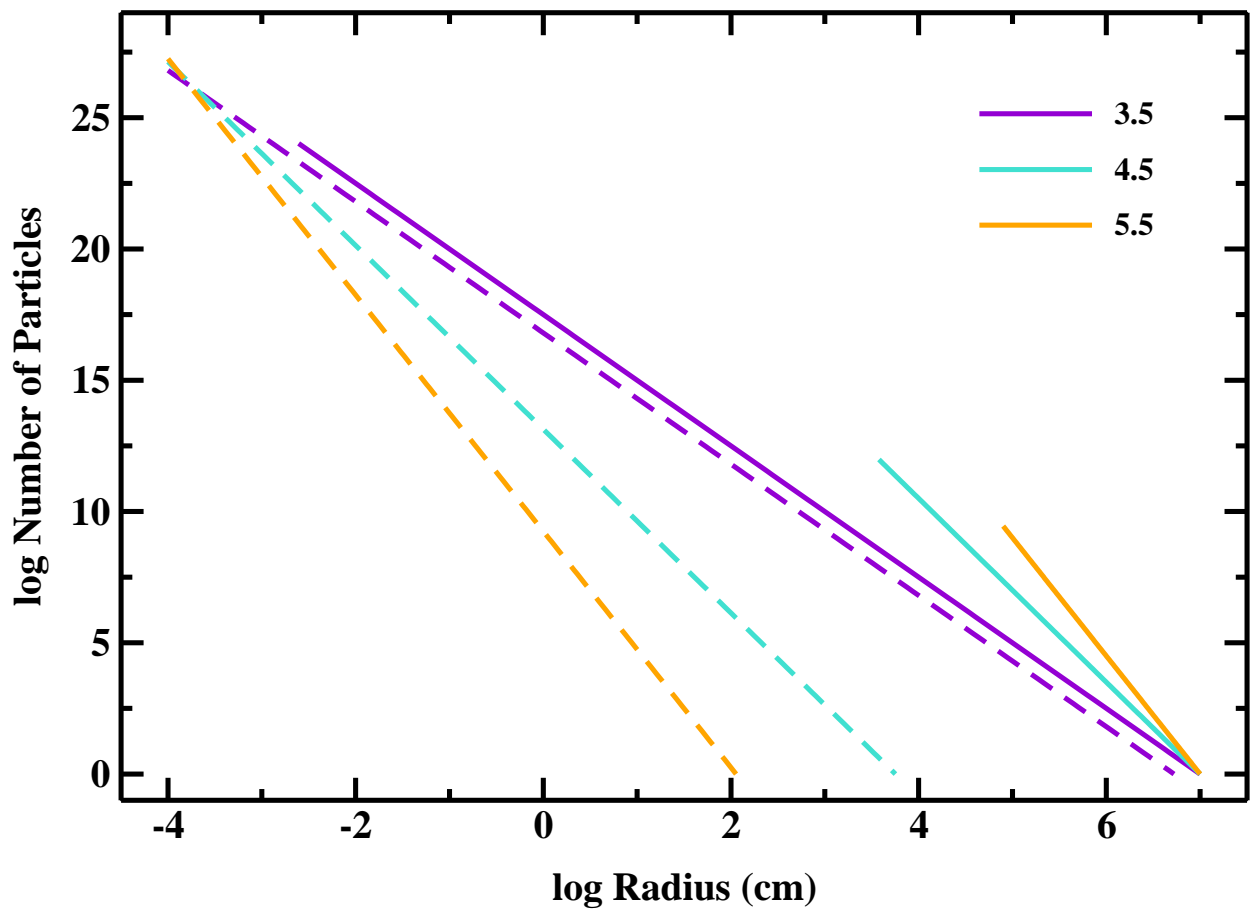}
\vskip 3ex
\caption{%
Size distributions for ensembles of particles with a total cross-sectional
area $\ad = 10^{20}$~cm$^2$. The legend indicates the slope $q$ of each
size distribution. 
Solid curves: size distributions requiring $n(R \ge \rmax) = 1$, with 
\rmax\ = 100~km. 
Dashed curves: size distributions requiring \rmin\ = 1~\mum.
Setting \rmin\ (\rmax) yields ensembles with fewer (more) large particles
and smaller (larger) total mass.
\label{fig: size-dist}
}
\end{figure}

To specify the size distribution completely, we set $q$ and {\it any two} of
\rmin, \rmax, or \ad. Fig.~\ref{fig: size-dist} shows an example where we 
adopt $\ad = 10^{20}$~cm$^2$ and either \rmin\ = 1~\mum\ (dashed curves) 
or \rmax\ = 100~km (solid curves) for $q$ = 3.5, 4.5, or 5.5.  When we fix 
\ad, $q$, and \rmin, the maximum size \rmax\ and the total mass \md\ follow 
from eqs. (\ref{eq: area}--\ref{eq: mass}).  Fixing \ad, $q$, and \rmax\ establish 
different values for \rmin\ and \md.  For any $q$, there are an infinite number 
of combinations of \rmin\ and \rmax\ that yield identical area \ad.  In general, 
setting a small value for \rmin\ (e.g., 1--10~\mum) leads to smaller \rmax\ and 
\md. Fixed \rmin\ also produces a small range in $N_d$ the total number of 
particles.  Setting a large value for \rmax\ results in larger \rmin\ and 
\md\ and a smaller $N_d$.

Specifying the size distribution in terms of \md\ is more complicated.
When $q < 4$ and $\rmax \gg \rmin$, setting $q$ and \md\ establishes
\rmax\ (eq. [\ref{eq: mass}]). Fixing \rmin\ then yields \ad. For
$q \ge 4$ (or when \rmax\ is not much larger than \rmin\ for any $q$), 
choosing {\it any two} of \rmin, \rmax, \ad, or \md\ then defines the 
remaining parameters. Although more cumbersome, this approach is an
integral part of planet formation theory. We will return to it when we 
consider specific models for dust in the next sections.

The main parameters of the size distribution -- \rmin, \rmax, and $q$ -- 
depend on physical events throughout the planet formation process.  In a 
collisional cascade, for example, the total mass in solids and the 
bulk properties of the solids establish $q$ and 
\rmax\ \citep[e.g.,][]{obrien2003,wyatt2008,kb2008,kriv2008,koba2010a,bely2011}.  
The luminosity of the central star sets \rmin; radiation pressure ejects 
smaller particles on short time scales compared to the local orbital 
period and the lifetime of the 
cascade \citep{burns1979}.  In a giant impact, the kinetic energy and the 
bulk properties of the protoplanets set \rmax, $q$, and the total mass of 
ejected material \citep[e.g.,][]{canup2004,canup2005,canup2011}. These 
quantities establish \rmin\ uniquely (eq. [\ref{eq: mass}]).

Within this framework, observations of the cross-sectional area of dust yield 
direct tests of planet formation theory. With the area known and \rmin\ derived
from the stellar luminosity, choosing $q$ then yields a unique \rmax.
Similarly, choosing \rmax\ implies a unique $q$. Once \rmin, \rmax, and
$q$ are known, comparisons with predictions from models of collisional 
cascades, giant impacts, or another mechanism provide clear tests of the 
theory.

\begin{figure}
\includegraphics[width=6.5in]{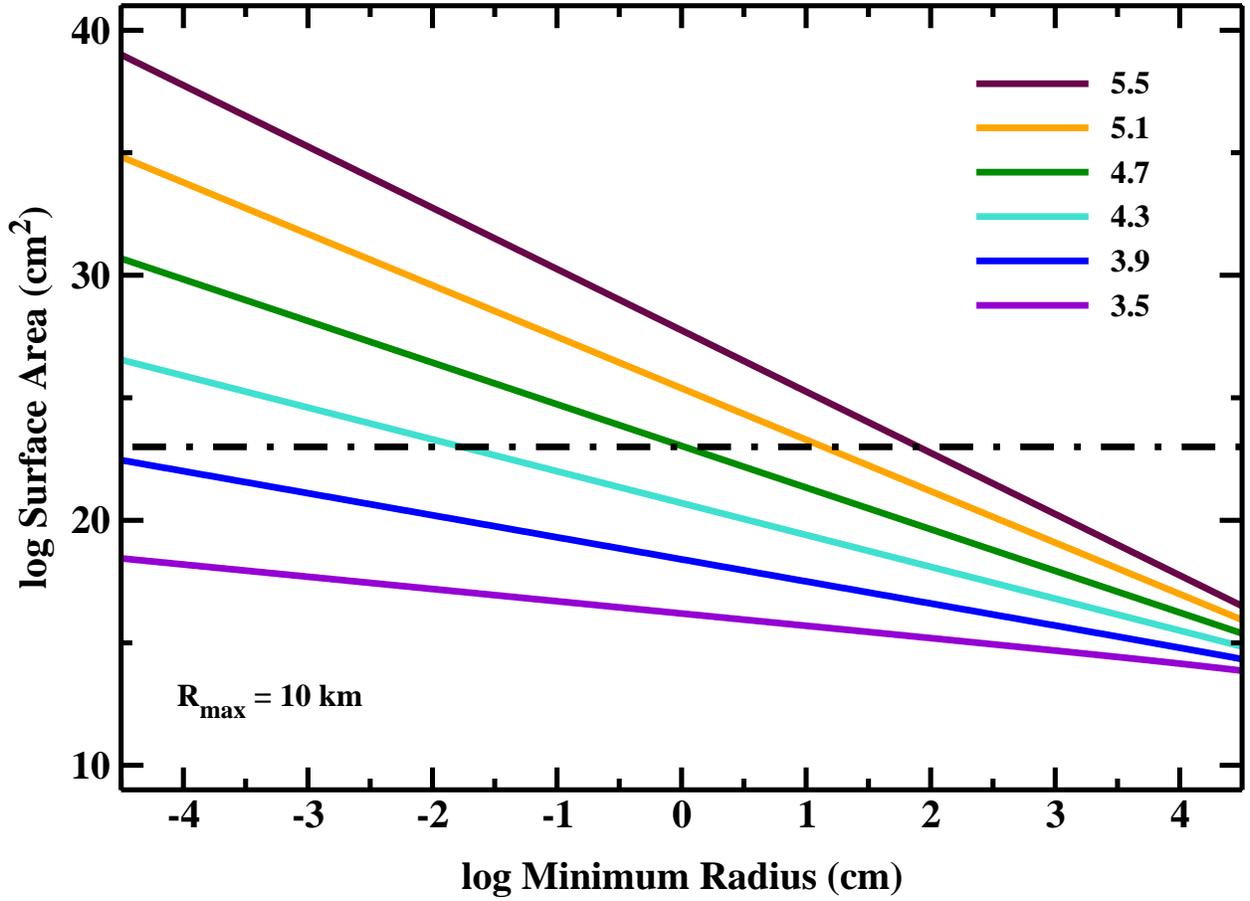}
\vskip 3ex
\caption{%
Relations between cross-sectional area and minimum radius for size distributions
with \rmax\ = 10~km and various $q$ as indicated in the legend. For each size 
distribution, $N(R \ge \rmax)$ = 1. The dot-dashed line indicates \ad\ =
$10^{23}$~cm$^2$.  At fixed \rmin, size distributions with larger \rmax\ and 
{\it smaller} $q$ have smaller area. 
\label{fig: area-rmin}
}
\end{figure}

To illustrate how these choices affect analyses of observations, we examine 
the variation of area with \rmin\ and \rmax. In Fig.~\ref{fig: area-rmin},
we set \rmax\ = 10~km, require one object with $R \ge \rmax$, and derive 
\ad\ as a function of \rmin\ and $q$.  The results behave as expected: 
ensembles of particles with larger \rmin\ have smaller area.  With 
\rmax\ and $q$ fixed, the cross-sectional area grows as $\rmin^{3-q}$.  
At fixed $q > 3$, increasing \rmin\ reduces the area. Similarly, the area 
grows with $q$ at fixed \rmin. As the size distribution becomes wider
or steeper, the area grows.

\begin{figure}
\includegraphics[width=6.5in]{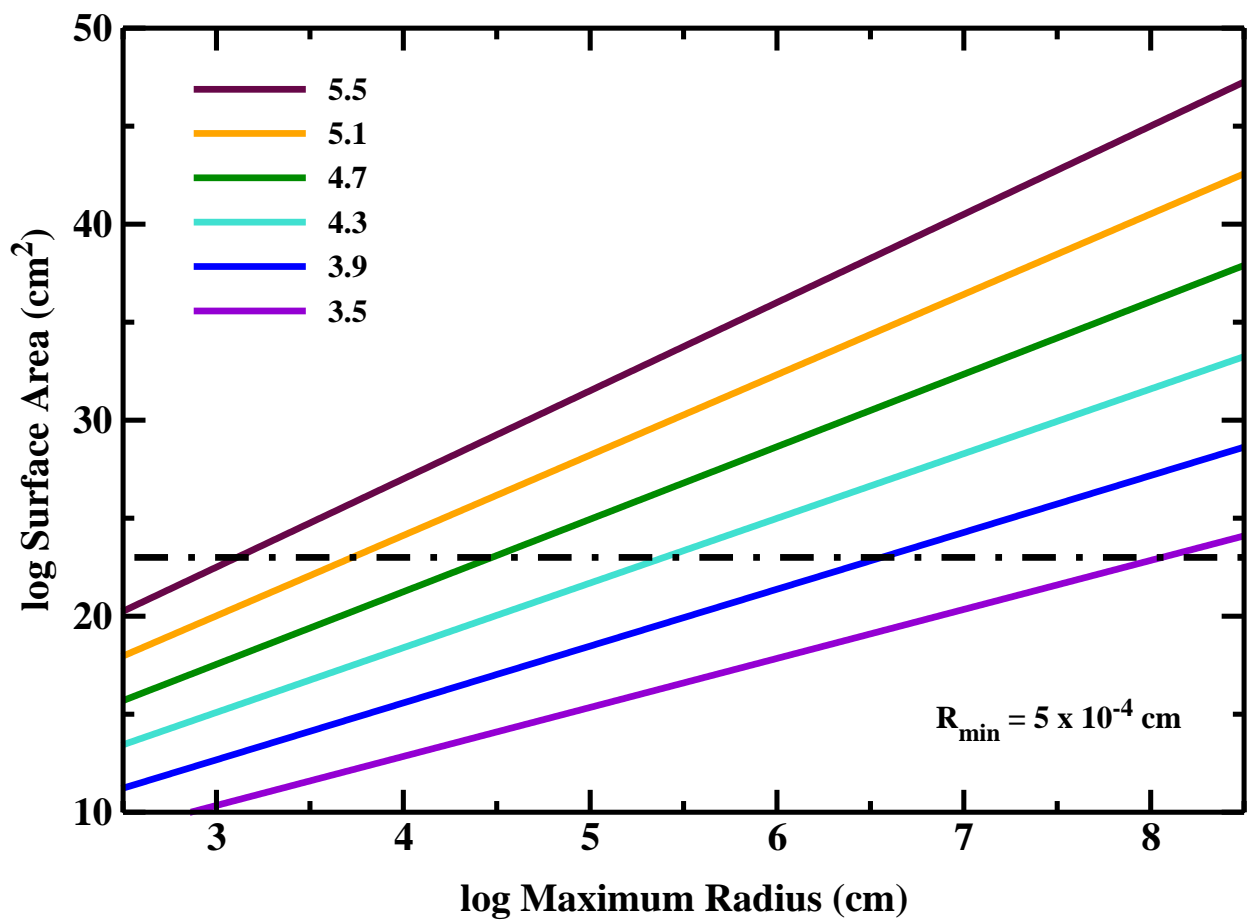}
\vskip 3ex
\caption{%
Relations between cross-sectional area and maximum radius for size distributions
with \rmin\ = 5~\mum\ and various $q$ as indicated in the legend. For each
size distribution, $N(R \ge \rmax)$ = 1.  The dot-dashed line indicates
\ad\ = $10^{23}$~cm$^2$.  At fixed \rmax, size distributions with larger 
\rmax\ and larger $q$ have larger area. 
\label{fig: area-rmax}
}
\end{figure}

With \rmin\ fixed, the area is also sensitive to \rmax\ (Fig.~\ref{fig: area-rmax}). 
Here, we set \rmin\ = 5~\mum, require one object with $R \ge \rmax$, and derive 
\ad\ as a function of \rmax\ and $q$.  At fixed \rmax, the area scales 
with $\rmax^{q-1}$ (eqs. [1--2]).  Thus, size distributions with larger $q$ 
have much larger area than those with smaller $q$.  Extending the size 
distribution to larger and larger \rmax\ yields larger \ad\ for all $q$.  
Although this result is somewhat counterintuitive, it is a consequence of our 
requirement of one object with $R \ge \rmax$.  The area grows with the number of 
very small objects, which grows as $\rmax^{q-1}$. Thus, for any $q > 1$, size 
distributions with larger \rmax\ have much larger area.  

For Fomalhaut b, these results place interesting constraints on models for dust
emission. The data reviewed in \S2 suggest an optically thin cloud with 
\ab\ $\approx 10^{23}$~cm$^2$ and \rmin\ $\approx$ 5~\mum. 
Fig.~\ref{fig: area-rmin} rules out size distributions with \rmax\ = 10~km and 
either $q \lesssim$ 3.9 or $q \gtrsim$ 4.1.  From Fig.~\ref{fig: area-rmax}, 
larger (smaller) \rmax\ yields more (less) area. Thus, optically thin models with 
$q \lesssim$ 3.9 can match the observed area with {\it larger} \rmax.  Similarly, 
optically thin models with $q \gtrsim$ 4.1 can match observations with 
{\it smaller} \rmax. 

Assuming Fomalhaut b has dust particles as small as 5--10~\mum, Fig.~\ref{fig: area-rmax} 
establishes combinations of $q$ and \rmax\ that match the observed area.
The implied range in \rmax\ is enormous: from $\rmax \approx$ 10~m for $q$ = 5.5 
to \rmax\ $\approx$ 1~km for $q$ = 4.5 to \rmax\ $\approx$ 1000~km for $q$ = 3.5.
From Fig.~\ref{fig: size-dist}, each of these prescriptions to achieve the target
\ad\ will have very different total masses. 

\begin{figure}
\includegraphics[width=6.5in]{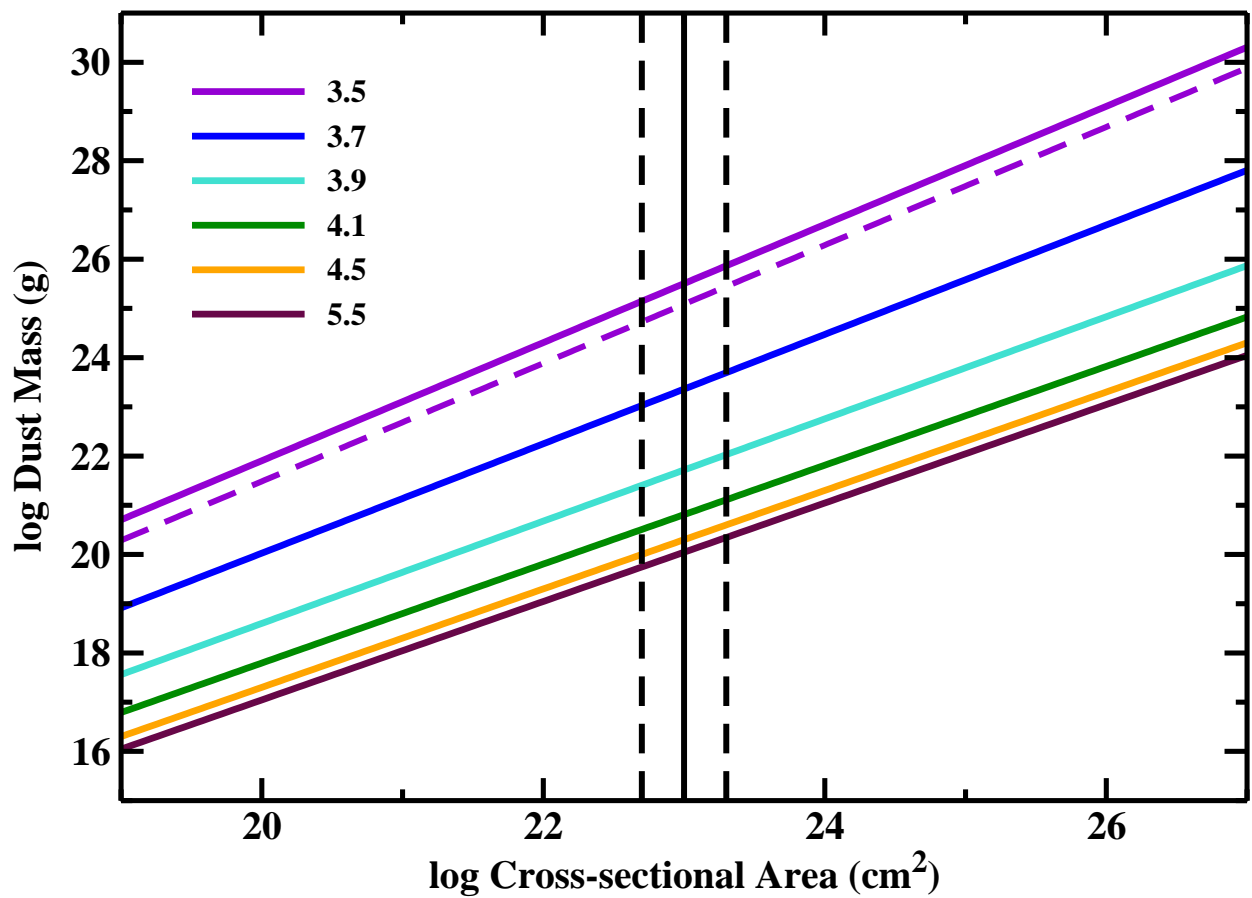}
\vskip 3ex
\caption{%
Relations between dust mass and cross-sectional area for power-law size
distributions (eq. [\ref{eq: size-dist}]), with \rmin\ = 5~\mum, a range of
\rmax\ (as in Fig.~\ref{fig: area-rmax}), and the range of $q$ indicated in 
the legend. The dashed curve repeats results for \rmin\ = 1~\mum.
Vertical lines: inferred cross-sectional area for a dust cloud in Fomalhaut b 
(solid line) with factor of two uncertainty (dashed lines). 
For fixed \ad, steeper size distributions (with larger $q$) require less total
mass in dust.
\label{fig: area-mdust}
}
\end{figure}

To establish limits on the total dust mass, Fig.~\ref{fig: area-mdust} plots 
\md\ as a function of \ad\ for \rmin\ = 5~\mum\ and various \rmax\ and $q$.  For 
fixed \ad, ensembles of dust with steeper size distributions (large $q$) require 
much less dust mass than ensembles with shallower size distributions (small $q$).
For fixed $q$, larger areas require larger masses. With $n_0 \propto \rmax^{q-1}$, 
the range in dust mass at small \ad\ is roughly two orders of magnitude smaller than 
at large \ad.  In Fomalhaut b, the range of likely dust masses is somewhat more 
than 5 orders of magnitude (\md\ = $10^{20}$~g for $q$ = 5.5 to \md\ = 
$2 \times 10^{25}$~g for $q$ = 3.5). 

To provide better constraints on the properties of the dust size distribution, we now 
consider plausible origins for the solid material in Fomalhaut b. After deriving 
constraints for dust produced in a giant impact, we explore the structure of a 
circumplanetary disk composed of (i) debris captured from the protoplanetary disk 
and (ii) debris from collisions of satellites orbiting the planet.

\subsection{Impact Models}

Giant impacts generally have two possible outcomes
(i) an expanding, isolated dust cloud orbiting the central star 
\citep[e.g.,][]{kb2005,galich2013} or
(ii) a disk or cloud of debris surrounding a (binary) planet
\citep[e.g.,][]{asphaug2006,canup2011,lein2012}.  The large area of the dust cloud in Fomalhaut b 
probably eliminates the second option.  At the distance of Fomalhaut b from Fomalhaut, 
likely giant impacts involve Earth-mass or smaller planets \citep{kb2008,kb2010}. 
Detailed SPH simulations \citep[e.g.,][]{canup2011} suggest most of the debris orbits 
the planet at less than 10--30 times the radius of the planet.  Although tidal forces
can expand the orbits of debris particles \citep{kb2014}, the likely outer radius of
the debris is still a factor of five to ten smaller than the minimum radius for a cloud 
in Fomalhaut b, $R_c \approx$ 300 \rearth\ \citep[\S2; see also][]{tamayo2013}.  
Thus, we explore models of debris within an isolated dust cloud.

\subsubsection{Ejected Mass}

We consider a simple head-on collision of two protoplanets with radii $R_1, R_2$, 
mass $M_1, M_2$, mass density $\rho_p$, and collision velocity $v_c$.  Defining
\mesc\ as the mass ejected from the event, the largest remnant has a mass
\mlr\ = $M_1 + M_2 - \mesc$.  If all the debris resides in a single object, 
$\mesc = 4 \pi \rho_p R_{esc}^3 / 3$.  The size of the largest object in the 
debris -- often called the largest fragment or the second largest remnant -- 
has $\rlf\ = \flf R_{esc}$ with $\flf \approx$ 0.1--0.8.  Thus, the mass 
of the largest fragment has a typical mass $ \mlf = f_{LF}^3 \mesc$  
$\approx 10^{-3} - 0.5 ~ \mesc$ 
\citep[see][and references therein]{benz1999,durda2004,giaco2007,lein2012}.

To estimate \mesc, we consider two prescriptions for high speed collisions in a
protoplanetary disk.  When $M_2 \ll M_1$, the impact produces a crater and ejects 
material from the surface of the larger protoplanet.  The ejecta have a power law 
distribution of velocities, with $f(v > v_c) \propto (v / v_c)^{-\alpha}$ and 
$\alpha \approx$ 1--3 \citep[e.g.,][and references therein]{gault1963,stoff1975,
okeefe1985,housen2003,housen2011}.  Although it is possible to derive the ejected mass 
from theoretical expressions for the kinetic energy of the impact and the binding energy 
of the larger protoplanet \citep[e.g.,][and references therein]{davis1985,lein2012}, 
\citet{housen2011} derive the ratio \mesc/$M_2$ from a variety of laboratory 
measurements of point-mass projectiles impacting much larger targets. Extrapolating
the results in their Fig. 16 suggests $\mesc / M_2 \approx f_{cr} (v_c / v_{esc})^\alpha$, 
where $f_{cr} \approx$ 0.01 and $\alpha \approx$ 1.0--1.5 \citep[for a recent application 
of this approach to asteroids in the Solar System, see][]{jewitt2012}.  For comparison, 
\citet{svet2011} derives $f_{cr} \approx$ 0.03 and $\alpha \approx$ 2.3 from a suite of 
theoretical calculations of cratering impacts \citep[e.g.,][]{green1978}.

To derive a simple expression for \mesc, we adopt $\alpha$ = 1.5 and a mass density
$\rho_p = 1.5$~g~cm$^{-3}$. Setting $v_{esc}$ as the escape velocity of the larger 
protoplanet \citep[e.g.,][]{jewitt2012,galich2013}:
\begin{equation}
{\mesc \over M_2} \approx 3.7 \times 10^4 f_{cr} v_c^{1.5} ~ R_1^{-1.5} ~ .
\label{eq: mesc-cr1}
\end{equation}
Larger impact velocities produce more debris. Impacts onto more massive protoplanets 
yield less debris.

For high velocity collisions between objects with roughly equal masses, results for 
cratering impacts provide a less accurate measure of the ejected mass 
\citep[e.g.,][]{davis1985,benz1999,asphaug2006,lein2009,lein2012}. Recent numerical 
simulations establish collision outcomes over a broad range in $M_2 / M_1$.  The 
ejected mass is fairly well-represented by a simple expression, 
\begin{equation}
{\mesc \over M_{tot}} \approx 0.5 \left ( { Q_c \over Q_d^* } \right )^\beta ~ .
\label{eq: mesc-qd}
\end{equation}
Here, $Q_c$ is the center-of-mass collision energy per unit mass, $Q_d^*$ is the collision 
energy per unit mass required to disperse 50\% of the total mass $M_{tot} = M_1 + M_2$ to 
infinity, and $\beta \approx$ 1--1.25.  The $Q_d^*$ term is roughly equivalent to the 
binding energy per unit mass and depends on the physical properties of the protoplanets: 
\begin{equation}
Q_d^* = Q_b R^{\epsilon_b} + Q_g \rho_p R^{\epsilon_g} ~ .
\label{eq: qdstar}
\end{equation}
In this expression, $R$ is the radius of a protoplanet with mass $M_{tot}$,
$Q_b R^{\epsilon_b}$ is the bulk component of the binding energy and 
$Q_g \rho_g R^{\epsilon_g}$ is the gravity component of the binding energy. For
most materials \citep[e.g.,][]{benz1999,lein2012}, the bulk (gravity) component of the 
binding energy dominates for solid objects with $R \lesssim$~0.01~km ($R \gtrsim$~0.01~km).

Here, we concentrate on catastrophic collisions of large objects in the gravity regime.  
For reduced mass $\mu = M_1 M_2 / (M_1 + M_2)$, $Q_c = 0.5 \mu v_c^2 / (M_1 + M_2)$.  
To compare with eq. (\ref{eq: mesc-cr1}), we adopt $M_2 \ll M_1$. The center of mass
collision energy is then $Q_c \approx 0.5 (M_2 / M_1) v_c^2$.  For icy objects with
$\rho_p = 1.5$~g~cm$^{-3}$, the binding energy parameters are $Q_g \approx 0.2$ and 
$\epsilon_g \approx 1.3$ \citep[e.g.,][]{benz1999,lein2008,lein2009}.  With 
$\beta \approx$ 1 \citep{lein2012}, the mass in debris is 
\begin{equation}
{\mesc \over M_2} \approx f_{cat,1} v_c^2 R^{-1.3} ~ ,
\label{eq: mesc-cr2}
\end{equation} 
where $f_{cat,1} \approx$ 0.83.

For equal mass protoplanets with $M_1 \approx M_2$, collisions have a center-of-mass
collision energy $Q_c = v_c^2 / 8$. Setting $R \approx \sqrt[3]{2} R_1$ for the radius 
of a merged object with $\mtot = M_1 + M_2 \approx 2 M_1 \approx 2 M_2$, the ejected 
mass is:
\begin{equation}
{ \mesc \over M_2 } \approx f_{cat,2} ~ v_c^2 ~ R_1^{-1.3} ~ ,
\label{eq: mesc-cata}
\end{equation}
with $f_{cat,2} \approx 0.31$.  Compared to collisions with $M_2 \ll M_1$, $Q_c$ is 
much smaller when $M_1 \approx M_2$. Thus, $f_{cat,2}$ is much smaller than $f_{cat,1}$.

To derive results for the ejected mass, we specify the collision velocity.  For two 
protoplanets on intersecting orbits around the central star, 
$v_c^2 = v_0^2 + v_{esc}^2$, where $v_0$ is the relative velocity of the two protoplanets
at infinity.  Small protoplanets have negligible self-gravity; the collision velocity 
is then the relative velocity. For large protoplanets with significant self-gravity
($R \gtrsim$ 10--100~km), the collision velocity is roughly the escape velocity.

\begin{figure}
\includegraphics[width=6.5in]{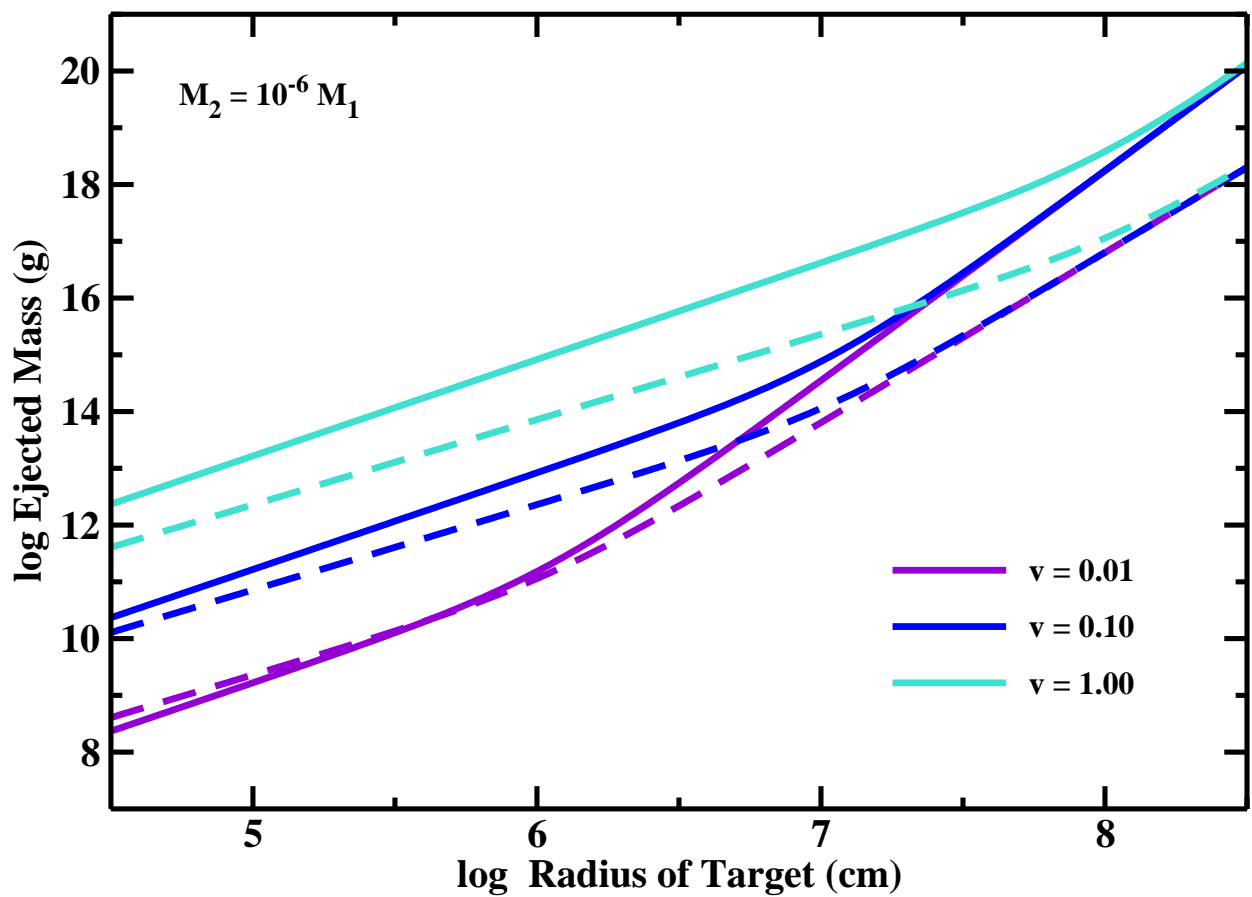}
\vskip 3ex
\caption{%
Relation between \mesc\ and $R_1$ for collisions with $M_2 = 10^{-6} M_1$ and various 
collision velocities in \kms. Solid curves: results using eq. (\ref{eq: mesc-cr2}). 
Dashed curves: results using eq. (\ref{eq: mesc-cr1}). In either approach, 
ejecting the minimum dust mass required in Fomalhaut b requires very high velocity 
collisions onto massive objects with $R_1 \gtrsim$ 1000~km.
\label{fig: mej-cr}
}
\end{figure}

Fig. \ref{fig: mej-cr} shows relations between \mesc\ and the radius of the target 
protoplanet for collisions with $M_2 = 10^{-6} M_1$ and several collision velocities. 
Low velocity impacts ($v_0 \approx$ 0.01-0.10~\kms) on low-mass targets 
($R \approx$ 1--10~km) yield little dust, \mesc\ $\approx 10^{10} - 10^{12}$~g. In 
this regime, the self-gravity of the larger protoplanet is negligible; \mesc\ depends 
only on $v_0$.  The two expressions for collisions with low mass projectiles then 
yield similar amounts of debris. 

As the impact velocity and target radius grow, the self-gravity of the protoplanet
becomes more and more important. The ejected mass is then independent of $v_0$ and
depends on the escape velocity of the larger protoplanet.  In this regime, the 
expression derived for cratering impacts (eq. [\ref{eq: mesc-cr1}]) yields much 
smaller amounts of ejected mass than results derived from fits to numerical simulations
(eq. [\ref{eq: mesc-qd}]).  Because their structure contains more flaws, larger objects 
are relatively easier to break than smaller objects \citep{benz1999,housen2003}. In 
higher velocity collisions, more of the target is involved in the collision. Higher 
velocity collisions onto larger targets then eject more material per unit collision 
energy. Fits to numerical simulations (eq. [\ref{eq: mesc-qd}]) capture this complexity 
more accurately than estimates derived from the escape velocity (eq. [\ref{eq: mesc-cr1}]).
Thus, the numerical results provide more accurate estimates for the ejected mass than
the analytic expression.

\begin{figure}
\includegraphics[width=6.5in]{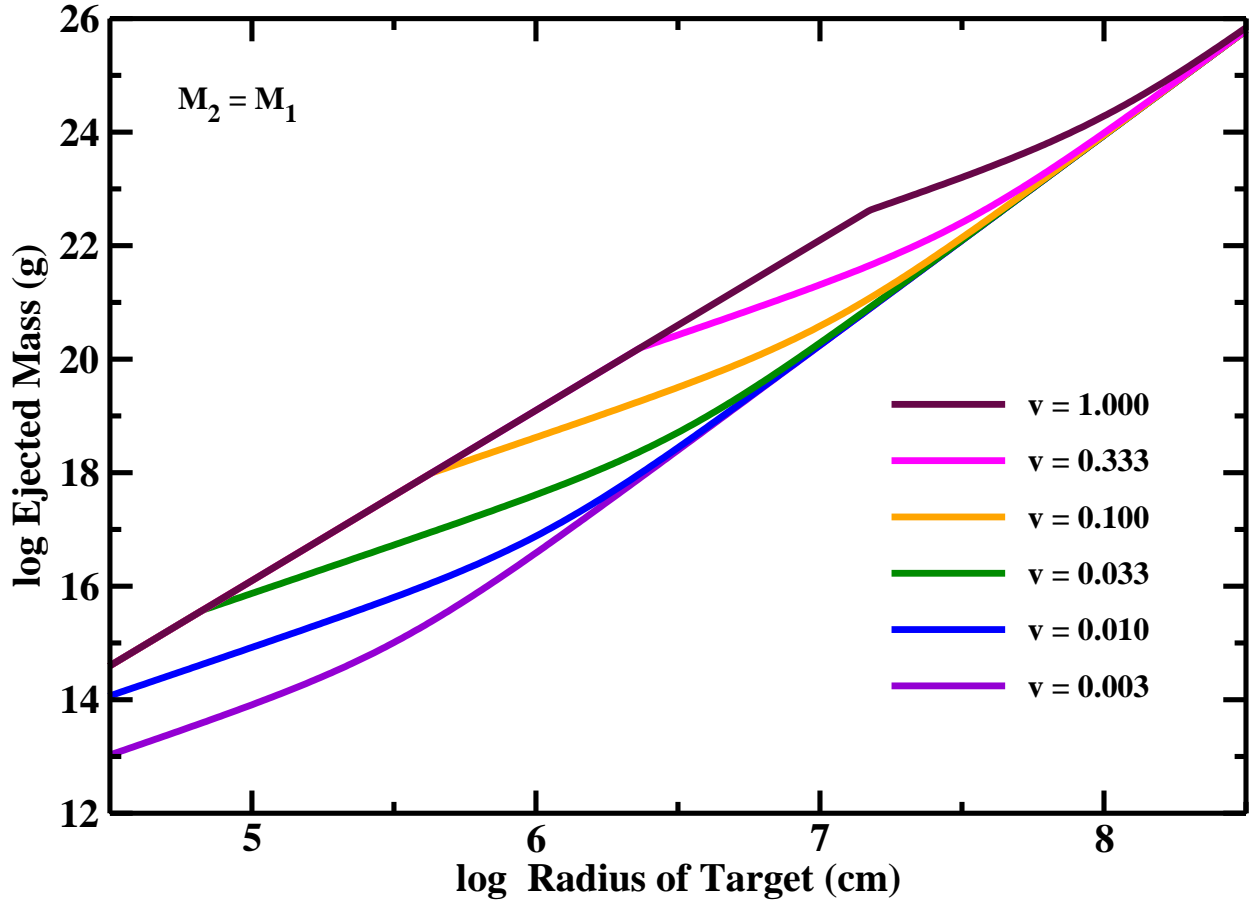}
\vskip 3ex
\caption{%
Relation between \mesc\ and $R_1$ for collisions with $M_2 = M_1$ and various 
collision velocities in \kms. Ejecting the broad range of plausible dust masses in
Fomalhaut b requires moderate to high velocity collisions between objects with 
radii of 10--1000~km.
\label{fig: mej-cat}
}
\end{figure}

Fig. \ref{fig: mej-cat} shows the relation between \mesc\ and radius for equal mass 
protoplanets.  When protoplanets are small, $R \lesssim$ 1--3~km, the highest velocity 
collisions completely disrupt the target. The ejected mass is then the sum of the two 
protoplanet masses, setting the upper left edge of the curves in Fig. \ref{fig: mej-cat}. 
For larger protoplanets, the escape velocity sets a lower limit on the impact velocity. 
This lower limit establishes the lower right edge of the curves. As a result, the range 
of ejected masses is fairly small -- roughly two orders of magnitude -- for any target 
radius.  Although larger protoplanets are less prone to complete disruption, collisions 
between Earth-mass protoplanets still eject several lunar masses of material 
\citep[e.g.,][]{canup2001,canup2004}. 

\subsubsection{Surface Area}

To derive the total cross-sectional area of fragments \ad\ from the ejected mass \mesc, 
we must specify the parameters of the size distribution. In laboratory experiments and 
theoretical simulations, $q$, \rmin, and \rmax\ depend on the parameters of the
experiment or the simulation \citep[e.g.,][]{housen2011,lein2012}. However, many of 
these quantities cannot be inferred from observations. Thus, we fix $q$ and derive 
\rmin\ and \rmax.  Setting $q < 4$ establishes the maximum radius, 
$\rmax\ \approx (3 (q-1) \mesc/ 4 \pi \rho)^{1/3}$ (eq. [\ref{eq: mass}]).  This approach 
yields a largest fragment with \flf\ = \rmax/\resc\ $\approx$ 0.6, which is close to
the sizes of the largest fragments observed in laboratory experiments or numerical 
simulations \citep{benz1999,lein2012}.  The minimum radius is formally arbitrary; for 
practical applications, stellar radiation pressure defines \rmin.

When $q > 4$ and $\rmax\ \gg$~1~cm, the first term in eq.~(\ref{eq: mass}) dominates.
To make progress, we consider a range of \flf\ = \rmax/\resc\ $\approx$ 0.01--0.6 which more 
than covers the typical range, \flf\ $\approx$ 0.1--0.6, in experiments and simulations.
Once \flf\ and \rmax\ are known, eq.~(\ref{eq: mass}) yields \rmin. 

\begin{figure}
\includegraphics[width=6.5in]{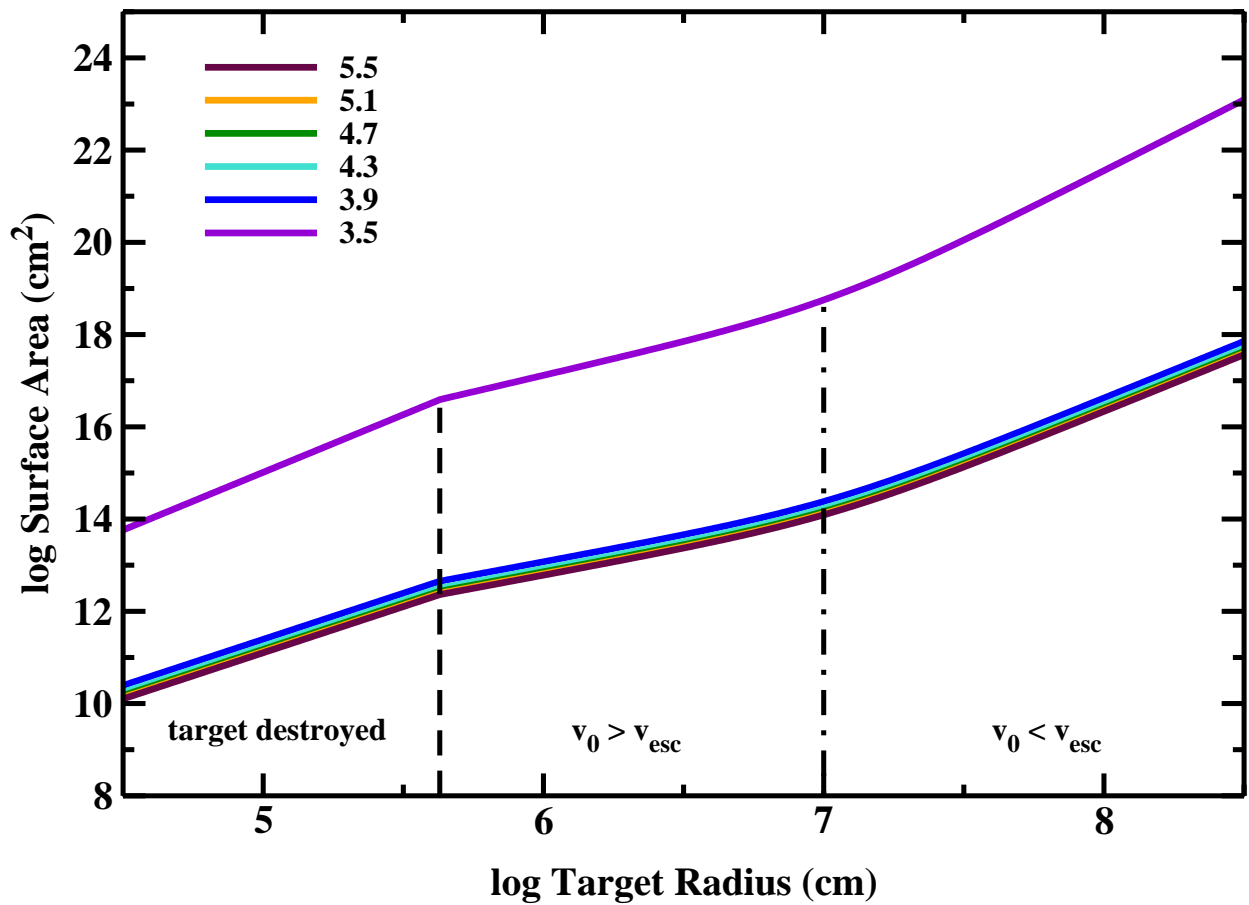}
\vskip 3ex
\caption{%
Surface area of ejecta as a function of the radius of the target for collisions 
between equal mass targets with $v_0$ = 0.1~\kms, \flr\ = 0.6, and various $q$ 
as listed in the legend. For large \flr, most of the mass is in the largest 
fragment. When the slope of the size distribution is relatively shallow 
($q \approx$ 3.5), the fragments extend to small sizes which have large 
surface area per unit mass. When the slope is steep, the ensemble of particles
has fewer low mass fragments and smaller surface area.
\label{fig: rad-area1}
}
\end{figure}

\begin{figure}
\includegraphics[width=6.5in]{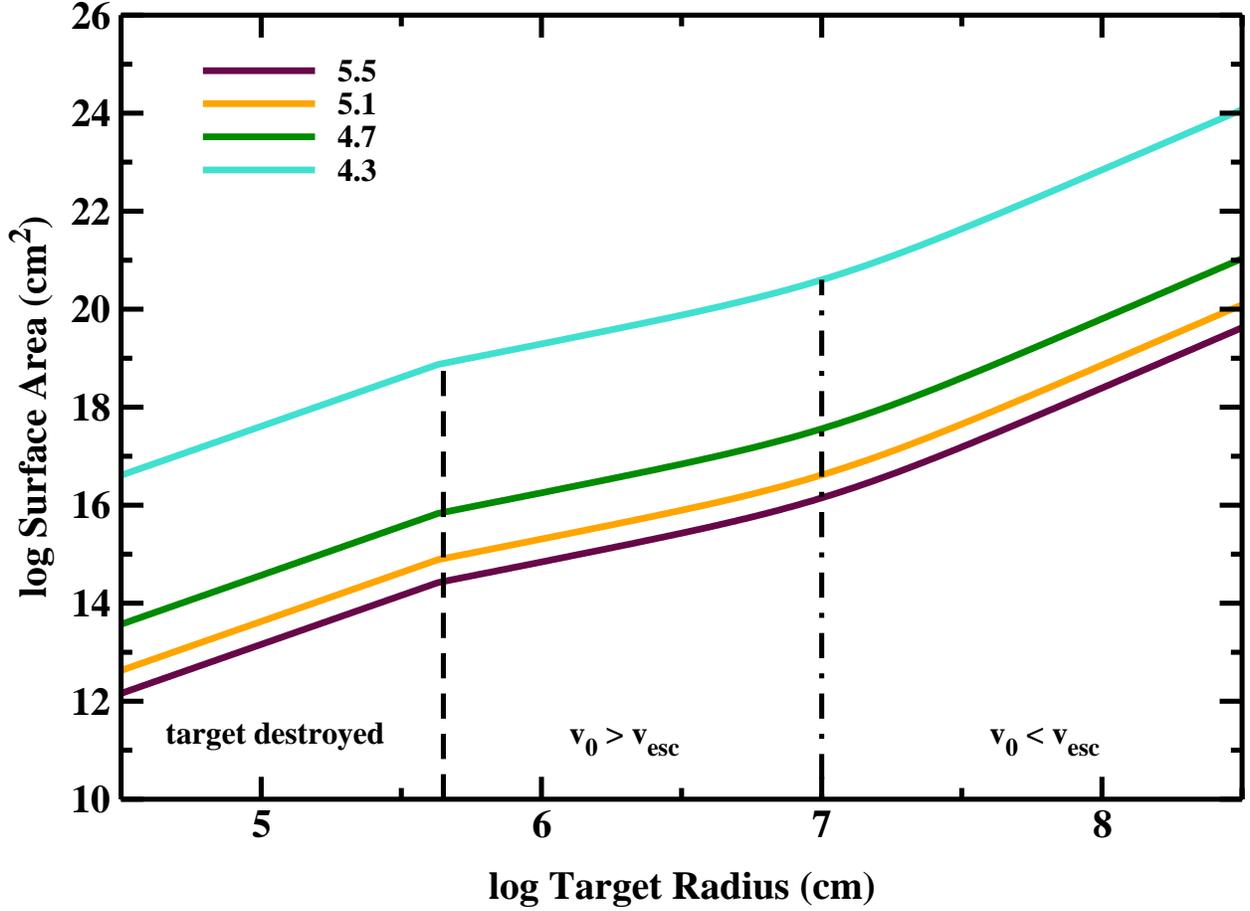}
\vskip 3ex
\caption{%
As in Fig.~\ref{fig: rad-area1} for \flr\ = 0.1. Placing less mass in the 
largest fragment leaves more mass for smaller fragments. Thus, smaller
\flr\ produces ensembles of dust with larger surface area per unit mass.
\label{fig: rad-area2}
}
\end{figure}

\begin{figure}
\includegraphics[width=6.5in]{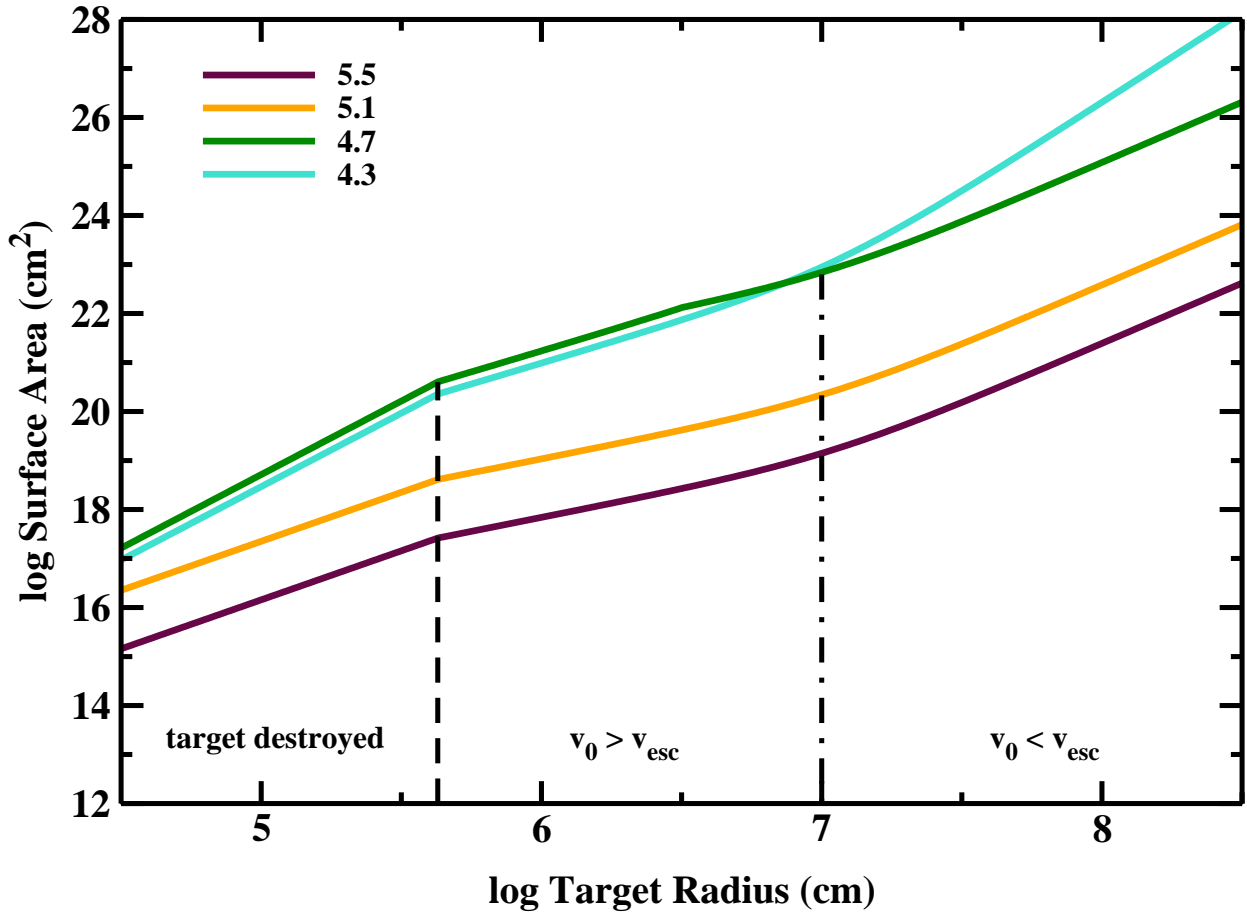}
\vskip 3ex
\caption{%
As in Fig.~\ref{fig: rad-area1} for \flr\ = 0.01. 
\label{fig: rad-area3}
}
\end{figure}

Figs.~\ref{fig: rad-area1}--\ref{fig: rad-area3} illustrates the variation of the area
with target radius for collisions between equal mass targets with $v_0$ = 0.1~\kms, various
$q$, and \flf\ = 0.6 (Fig.~\ref{fig: rad-area1}), \flf\ = 0.1 (Fig.~\ref{fig: rad-area2}), and
\flf\ = 0.01 (Fig.~\ref{fig: rad-area3}). For $q$ = 3.5 and 3.9, requiring one object with
a radius of \rmax\ establishes \flf\ = 0.6; thus, only Fig.~\ref{fig: rad-area1} shows results 
for these values of $q$.

For each value of $q$ (indicated by the legend), the relation between the cross-sectional
area and target radius has three regimes. Small protoplanets with $R \lesssim$ 5~km are
the weakest and the easiest to break. In collisions with modest velocities, the 
projectile and the target are completely destroyed. The area of the 
ejected material then increases with target radius as $\ad \propto R^{2.5}$. For 
somewhat larger protoplanets (indicated by the vertical dashed line in each Figure), 
the binding energy per unit mass grows slowly ($\propto R^{1.3}$) compared to an
ideal monolithic object ($\propto R^2$). Modest velocity impacts do not destroy 
these protoplanets. When $v_0 \gtrsim v_{esc}$, the area of the ejecta grows 
slowly with increasing radius, $\ad \propto R^{1.5}$. Among the largest protoplanets, 
where $v_0 \lesssim v_{esc}$, collisions occur at the escape velocity. The collision energy
then scales with $v_{esc}^2 \propto R^2$, which grows much faster with radius than the 
binding energy ($\propto R^{1.3}$). The area then grows roughly with the volume 
of the protoplanets, $\ad \propto R^3$. 

The variation of cross-sectional area with $q$ has a different topology in each Figure.  
When \flf\ = 0.6 (Fig.~\ref{fig: rad-area1}),
roughly 20\% of the ejected mass lies in the largest fragment. For steep
size distributions with $q \gtrsim$ 3.9, the number of smaller particles increases very 
rapidly with radius; \rmin\ is always large, roughly a few meters to several tens of 
meters. The ratio of the area to the mass is then small. Thus, ensembles of particles
with \flf\ = 0.6 and $q \gtrsim$ 3.9 have small area. For more shallow size 
distributions with $q \approx$ 3.5, the size distribution extends to much smaller radii,
\rmin\ $\approx$~1~\mum. These ensembles have much larger area.

When \flf\ = 0.1 (Fig.~\ref{fig: rad-area2}), ensembles with $q \gtrsim$ 3.9 have much 
larger area.  As \flf\ declines, the largest fragment has a smaller and smaller
fraction of the total mass. With more mass available for smaller objects, the size
distribution extends to smaller \rmin. Ensembles of particles with smaller \rmin\ have
larger area (eq. [\ref{eq: area}]). 

This trend continues for \flf\ = 0.01 (Fig.~\ref{fig: rad-area3}). When the largest 
fragment has only 0.0001\% of the total ejected mass, the size distribution can
extend to the smallest allowed sizes (\rmin\ = 5~\mum\ for Fomalhaut). For 
$q \approx$ 4.3--4.7, the cross-sectional area saturates for target radii smaller than 
roughly 100~km. For larger $q$, \rmin\ is much larger than 5~\mum; the area per unit 
ejected mass then remains fairly small.

For the observed cross-sectional area of roughly $10^{23}$~cm$^2$ in Fomalhaut b, these 
results provide clear constraints on plausible protoplanets involved in a single giant 
impact.  Adopting the \flf\ $\approx$ 0.1--0.6 derived from numerous theoretical 
simulations sets a firm upper limit on the radius of the target, $R \lesssim$ 1000--2000~km.  
Extending the plausible range of fragment sizes to \flf\ $\approx$ 0.01 allows collisions 
among smaller targets, $R \sim$ 100~km, providing $q \approx$ 4.3--4.7.

\subsubsection{Summary}

Using only collision dynamics and the properties of power-law size distributions, we
generate several useful expressions for the mass ejected during a collision of two high 
velocity objects. In all collisions, the ejected mass depends on the escape velocity 
and the relative velocity of the impactors.  When the mass ratio between the impactors 
is large (Fig. \ref{fig: mej-cr}), large ejected masses require high velocity collisions 
onto very massive protoplanets.  When the mass ratio is near unity, somewhat less energetic 
collisions yield comparable amounts of ejected material (Fig. \ref{fig: mej-cat}).  For 
the relative velocity expected during the late stages of planet formation, the range in 
the ejected mass is 2--3 orders of magnitude (Fig. \ref{fig: mej-cat}). Coupled with our 
expressions for the area (eq. [\ref{eq: area}]) and mass (eq. [\ref{eq: mass}]), these 
results yield the cross-sectional area as functions of the radii of the impactors and 
the fraction of mass \flf\ in the largest fragment of the debris 
(Figs.~\ref{fig: rad-area1}--\ref{fig: rad-area3}).
For \flf\ = 0.6, ensembles of particles with $q \gtrsim$ 3.9 have little area per unit 
mass. As \flf\ in the ejecta decreases, particles with steeper size distributions have 
larger area.

Applying this analysis to Fomalhaut b strongly favors impacts between roughly equal
mass protoplanets.  For ensembles of particles in an extended dust cloud containing 
one large object with $R = \rmax$ and no small objects with $R \lesssim \rmin$ = 5~\mum, 
we set limits on \rmax\ and the total mass \md\ as functions of $q$.  Our results indicate
$\rmax \approx$ 10~m and \md\ $\approx 10^{20}$~g for $q$ = 5.5, 
$\rmax \approx$ 1~km and \md\ $\approx 10^{20}$~g for $q$ = 4.5, 
$\rmax \approx$ 30~km and \md\ $\approx 2 \times 10^{23}$~g for $q$ = 3.9, and 
$\rmax \approx$ 1000~km and \md\ $\approx 3 \times 10^{25}$~g for $q$ = 3.5. 

Quantitative models for the ejected mass as a function of the collision energy place 
additional limits on the giant impact picture. Standard results for the radius of the 
largest fragment in a high velocity collision suggest target radii of 1000--2000~km. 
If laboratory experiments and numerical simulations overestimate the typical size of 
the largest fragment by a factor of ten, collisions between two 100~km protoplanets
produce enough dust when $q \approx$ 4.3--4.7.

These results limit the practicality of giant impact models for dust in Fomalhaut b
\citep[e.g.,][]{kb2005}.
When $f_{LF} \gtrsim$ 0.1, the required impactors are very large with radii of
1000--2000~km. Collisions between such large objects are very rare. For the surface
density at 30--130~AU outlined in \S2.1 (eq. [\ref{eq: sigma}]), a lower limit on
the collision rate for two 1000~km objects within a 10~AU annulus is 1 per 50--100~Myr
\citep[][Appendix]{kb2008}.  With a typical cloud lifetime of a few orbits or less 
\citep{kb2005}, detecting dust from this collision is very unlikely.

Allowing $f_{LF} \approx$ 0.01 allows smaller targets with radii of 100~km.  Collisions
between two 100~km objects within a 10~AU annulus centered at 120~AU are fairly common, 
with a lower limit of roughly once every $5 \times 10^3 - 10^4$ yr.  Although numerical 
simulations of collisions between pairs of 100~km particles often yield debris with 
$q \approx$ 4.5 \citep{lein2012}, outcomes with $f_{LF} \approx$ 0.01 are rare.  Despite 
the modest frequency, collisions which yield such small fragments seem unlikely.

\subsection{Continuous Capture into a Circumplanetary Cloud}

Originally envisioned as an explanation for the origin of the Moon \citep{ruskol1961},
the capture of circumstellar material onto circumplanetary orbits provides an interesting
alternative to dust formation from a giant impact 
\citep[for an application to satellite formation around Jupiter, see][]{estrada2006,koch2011}. 
In the simplest form of this model, an object enters the Hill sphere of a planet and 
collides with another object passing through the Hill sphere \citep{ruskol1972}, a
satellite of the planet \citep{durda2000,stern2009}, or the planet \citep{wyatt2002,kw2011a}.  
Close approaches between a low-mass binary and a planet 
\citep[][and references therein]{agnor2006} or two planets \citep{nesv2007} often yield a 
bound satellite.  Sometimes dynamical interactions with objects outside the Hill sphere 
produce a bound satellite \citep{ruskol1972,gold2002}.  In a variant of this mechanism, 
objects find temporary orbits around the planet and become bound after collisions with 
other small objects or dynamical interactions with other planets 
\citep{kort2005,suet2011,pires2012,suet2013}.

\subsubsection{Captured Mass}

To make an initial exploration of this picture for the formation of dust clouds 
surrounding an exoplanet, we estimate the capture rate from collisions of two
circumstellar objects within the Hill sphere of a planet\footnote{\citet{kw2011a} 
consider dust production from impacts with the planet.}. In this mechanism, material
enters the Hill sphere at a rate $\dot{M}_H$. The probability of a collision in the
Hill sphere is the optical depth $\tau_c$ of the circumstellar disk in the vicinity 
of the planet. After the collision, the planet captures a fraction $f_{cap}$ of the
material into bound orbits. The capture rate is then 
$\dot{M}_{cap} \approx \dot{M}_H \tau_c f_{cap}$.

The rate $\dot{M}_H$ depends on $\Sigma$ the local surface density of material, 
$\sigma$ the cross-section of the Hill sphere, and $\Omega$ the local angular frequency 
of the planet's orbit \citep[e.g.,][]{liss1987,gold2004}. For objects with a modest 
amount of gravitational focusing, $\dot{M}_H \approx 3 \Sigma \sigma \Omega$. We adopt 
a power-law surface density with the parameters from \S2.1.  For a planet with mass $M_p$ 
around a star of mass \mstar, the cross-section of the Hill sphere is $\pi \rhill^2$.
Material outside $a \gtrsim \gamma \rhill$ with $\gamma \approx$ 0.3--0.4 is unbound 
\citep[e.g.,][]{ham1992,ham1997,toth1999,shen2008,martin2011}. 
Thus, $\sigma \approx \pi \gamma^2 \rhill^2$.  

The optical depth depends on the size distribution of circumstellar objects. During the
late stages of the planet formation process, large objects contain nearly all of the mass 
and have a roughly power-law size distribution \citep[e.g.,][]{weth1989,koba2010a,kb2012}.  
To set plausible limits on the optical depth in this regime, we consider two approaches.
To establish a reasonable lower limit on the optical depth, we adopt a mono-disperse 
set of objects with radius \rmaxd\ and mass density $\rho_d$ = 1~\gcmc; then
$\tau_{c,m} \approx 3 \Sigma / 4 \rmaxd $.
For a reasonable upper limit, a size distribution (eq. [\ref{eq: size-dist}]) 
with \rmin\ = 1~km and $q \approx$ 4 yields $\tau_{c,sd} \approx 20 \tau_{c,m}$. 
We set \rmaxd\ = 100~km for both limits.

At 100 AU, the typical optical depth is small. With \rmaxd\ = 100~km and 
$\Sigma \approx$ 0.3~\gcms, $\tau_{c,m} \approx 3 \times 10^{-8}$. To match the
observed lower limit of $\tau_b \gtrsim 10^{-3}$, planets must capture at least 
$10^5$ times the amount of material passing through their Hill spheres.

The fraction of colliding material captured by the planet depends on the relative velocities
of planetesimals and the escape velocity of the planet \citep[e.g.,][]{ruskol1972,weiden2002}.
For material at 0.2--0.3~\rhill, $f_{cap} \approx 1 - 3 \times 10^{-3}$. Thus, the planet 
captures less than 1\% of material colliding within its Hill sphere. 

Combining $\dot{M}_H$, the two limits for $\tau$, and $f_{cap}$, the total capture rate is
\begin{eqnarray*}
\dot{M}_{cap} & \approx &
4.25 - 85 ~ \times ~ 10^{13} 
\left ( { f_{cap} \over 2 \times 10^{-3} } \right )
\left ( { d \over 1.0 } \right)^2 
\left ( { \gamma \over 0.3 } \right )^2
\left ( { \Sigma_0 \over {\rm 30~g~cm^{-2}} } \right )^2
\left ( { M_p \over \mearth } \right )^{2/3} \\
& &~~~~~~~~~~~~~~~~~~~~~~ \left ( { r \over {\rm 120~AU} } \right )^{-3/2}
\left ( { \mstar \over 2 ~ \msun } \right )^{11/6}
\left ( { \rmaxd \over {\rm 100~km} } \right )^{-1} ~ { \rm g~yr^{-1} } ~ . ~~~~~~~~~~~ (15) \\
\label{eq: mdot-cap}
\end{eqnarray*}
During a 100~Myr time frame in regions where $d \approx$ 1, an Earth-mass planet 
captures roughly $M_{cap} \approx 0.5 - 10 \times 10^{22}$~g of solid material 
into orbits with $a \approx$ 0.2--0.3~\rhill. More massive planets capture more 
material from the circumstellar disk.

Current observations of the Fomalhaut debris disk place useful constraints on $M_{cap}$.
As we outlined in \S2, the belt of dust at 130--155 AU contains roughly twice as much 
dust as the region from 35--130~AU \citep[e.g.,][]{acke2012}. The surface density of 
solids in the main belt is then six times larger than the surface density of solids 
in the inner disk.  A planet orbiting within the belt captures material roughly 50 
times faster than a planet orbiting at 30--130~AU.

\addtocounter{equation}{1}

\subsubsection{Size Distribution and Evolution of Captured Material}

Producing the observed cross-sectional area from captured material requires a 
size distribution dominated by small objects. If \ad\ $\approx 10^{23}$~cm$^2$ 
and $M_{cap} \approx 10^{23}$~g, the typical particle size is 0.5--1~cm.  
The minimum radius for captured particles depends 
on the ratio of the radiation force from the star to the gravity of the planet 
\citep{burns1979}. For a 1-10~\mearth\ planet at $r$ = 120~AU, the minimum stable 
radius for a single particle is 100--300~\mum\ \citep{burns1979,kw2011a}.  These
particles are ejected on time scales comparable to the orbital period of the 
planet around the central star.  During this time, the particles make several
circumplanetary orbits and may collide with other particles within the Hill sphere 
of the planet.  Particles with much smaller sizes are ejected on the local dynamical 
time scale and unlikely to interact with particles on stable orbits. 

\begin{figure}
\includegraphics[width=6.5in]{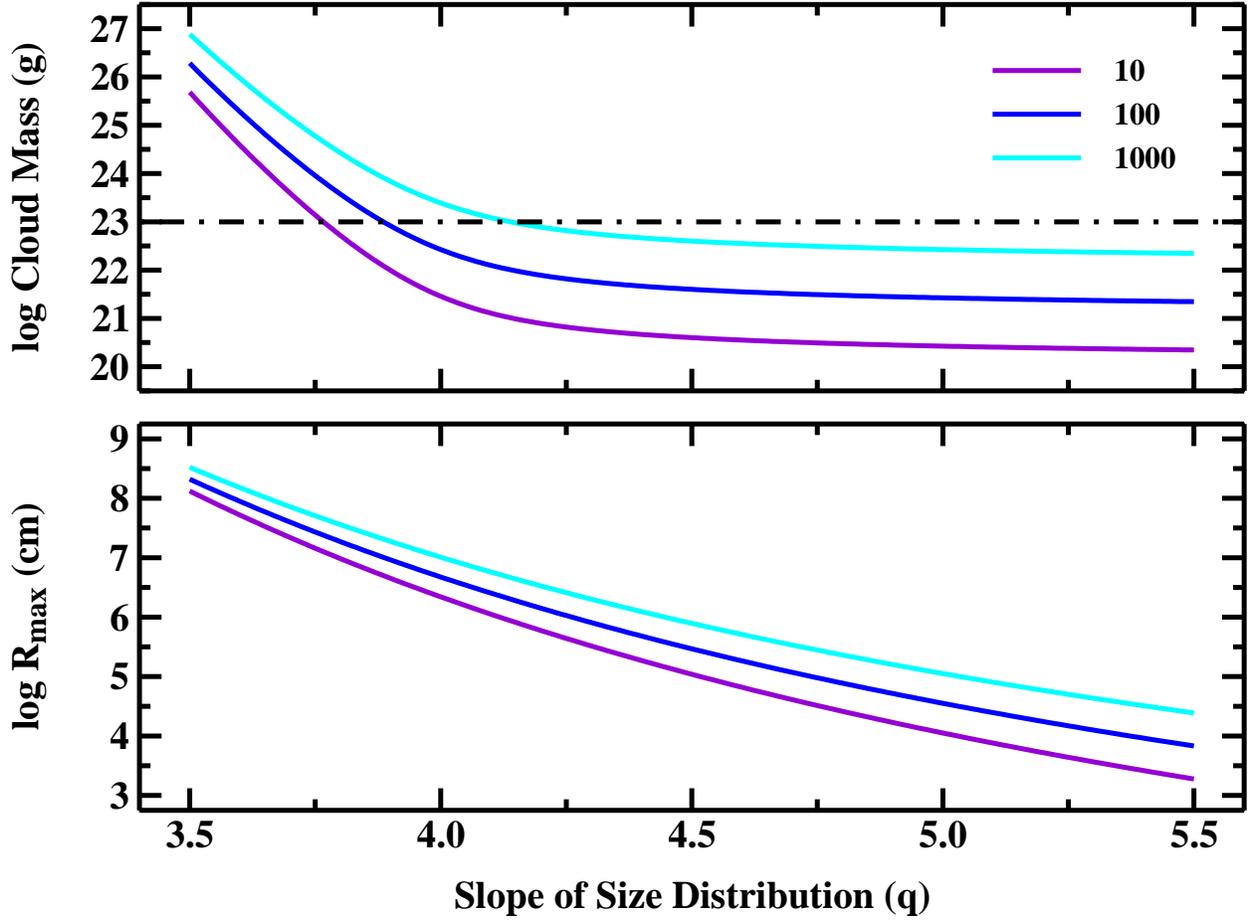}
\vskip 3ex
\caption{%
Size of the largest object \rmax\ (lower panel) and cloud mass \md\ (upper panel) as 
a function of the slope of the size distribution $q$ for \ad\ = $10^{23}$~cm$^2$ and 
various \rmin\ (in \mum) as indicated in the legend.  The horizontal dot-dashed line 
in the upper panel shows the maximum possible cloud mass from the capture model 
(eq. [\ref{eq: mdot-cap}]).  Matching the observed \ab\ with \md\ $\lesssim 10^{23}$~g
requires $q \gtrsim$ 4 and \rmax\ $\lesssim$ 50--60~km.
\label{fig: capt-mass}
}
\end{figure}

To derive combinations of \rmax\ and $q$ where $\ad = \ab \approx$ $10^{23}$~cm$^2$ 
with $\md\ \lesssim 10^{23}$~g, we set \rmin\ = 10--1000~\mum\ and calculate \rmax\ and 
\md\ as a function of $q$ (Fig.~\ref{fig: capt-mass}). For clouds with $q \lesssim$ 3.9, 
the mass required to match the observed \ab\ exceeds the likely maximum amount of 
captured material, $\sim 10^{23}$~g.  These models fail.  When $q \approx$ 4, clouds 
with \rmax\ $\approx$ 30--50~km and $\md\ \approx 10^{22} - 10^{23}$~g yield the 
observed area. Clouds with $q \approx$ 4.5--5.5 have \rmax\ $\approx$ 0.1--1~km and 
total masses $\md\ \approx 10^{21}$~g. 

As the cross-sectional area of a cloud approaches $\ad \approx 10^{23}$~cm$^2$, 
the long-term evolution depends 
on collision outcomes, collision rates, and the total angular momentum.  Captured 
fragments typically have semimajor axis $a_f \approx \gamma \rhill$ and large 
eccentricity $e_f \gtrsim$ 0.3. If the distribution of inclination angles relative 
to the plane of the circumstellar disk is random, each fragment has a randomly oriented 
angular momentum vector with specific angular momentum 
$L_f \approx (G m a_f (1 - e_f^2))^{1/2}$ \citep[e.g.,][]{dones1993}. On average, 
the total angular momentum is zero with a standard deviation of roughly $\sqrt{N} L_f$. 
If the planet captures material with somewhat higher or lower specific angular momentum 
than the planet, captured material may have a significant total angular momentum
\citep{dones1993}.

For captured particles with a range of radii, collision outcomes are sensitive to 
particle size.  Particles in a roughly spherical cloud have typical collision velocity 
$v \approx 1.3 (G M_p /a_f)^{1/2}$ 
\citep[e.g.,][and references therein]{kw2011a}.  Collisions with large kinetic energy 
relative to the binding energy produce debris; small collision energies allow mergers. 
For R $\gtrsim$ 0.01~km, the binding energy grows rapidly with radius. Thus, collisions
add mass to large particles and remove mass from small particles.  To identify the 
boundary between these regimes for collisions between unequal mass particles, we set 
$M_{esc} / M_2 \gtrsim 1$ in 
eq.~(\ref{eq: mesc-cr1}):
\begin{equation}
R_{d,u} \lesssim 7 \left ( { M_p \over 1~\mearth\ } \right )^{1/3} ~ .
\label{eq: rd1}
\end{equation}
For collisions among equal mass particles, setting $M_{esc} / M_{tot} \lesssim$ 0.1
in eq. (\ref{eq: mesc-cata}) yields a similar relation:
\begin{equation}
R_{d,e} \lesssim 5 \left ( { M_p \over 1~\mearth\ } \right )^{1/2} ~ .
\label{eq: rd2}
\end{equation}
For planets with $M_p \approx$ 0.1--10~\mearth, objects with $R \gtrsim$ 5--10~km 
grow slowly with time.  Collisions destroy all smaller particles.

Particle sizes also set the collision rates.  The typical lifetime of a 
100~\mum\ particle is short 
\begin{equation}
t_s \approx 5 \times 10^3 \left ( { 10^{23}~{\rm cm^2} \over \ad } \right ) ~ {\rm yr} ~ .
\label{eq: t-small}
\end{equation}
On this time scale, collisions convert 100~\mum\ particles into much smaller particles
which are unstable to radiation pressure. These collisions reduce the mass and 
cross-sectional area of the cloud.

For 10~km objects, the typical lifetime is much longer, 
\begin{equation}
t_l \approx 5 \times 10^6 \left ( { 10^{23}~{\rm cm^2} \over \ad } \right ) ~ {\rm yr} ~ .
\label{eq: t-large}
\end{equation}
Throughout their lifetimes, these large objects continually replenish the supply of 
much smaller objects. Although the cloud mass remains roughly constant, collisions 
among large objects increase the cross-sectional area of the cloud.

For the nominal capture rate in eq. (\ref{eq: mdot-cap}), the typical lifetime of
10~km objects implies a low mass cloud with a steep size distribution.  If captures
replenish the cloud on a $ 5 \times 10^6$ yr time scale, the cloud has a typical 
mass $\md \approx 5 \times 10^{21}$~g. Larger (smaller) capture rates allow 
a larger (smaller) cloud mass. To match the nominal \md\ and \ad\ for \rmin\ = 
100~\mum, the size distribution has \rmax\ $\lesssim$ 100~km and $ q \gtrsim$ 4.

Although factor of ten changes to \rmin\ have little impact on our conclusions
(Fig.~\ref{fig: capt-mass}), changing the capture rate allows a broader range of 
possible matches to observations. Larger (smaller) capture rates imply shallower 
(steeper) size distributions with larger (smaller) \rmax. Thus, matching the 
observed \ab\ with factor of 10--1000 increases in the capture rate is possible 
with $q =$ 3.9--3.5.  However, reducing the capture rate by a factor of 10 or
more eliminates all power-law size distributions with $\rmin \gtrsim$ 30~\mum.
In these situations, the cloud mass is too small to match the observed \ab.

\subsubsection{Summary}

This discussion establishes an evolutionary sequence for a capture model in
Fomalhaut b.  We envision a long series of protoplanet collisions within the 
Hill sphere of a much larger planet. These collisions gradually produce a cloud 
of satellites orbiting the planet, with sizes ranging from \rmin\ $\sim$ 
100~\mum\ up to \rmax\ $\approx$ 10--20~km. As the mass of the cloud grows,
collisions among captured objects eventually produce a collisional cascade
where objects with $R \lesssim$ 5~km are slowly ground into smaller and
smaller objects. Continuous captures from the circumstellar disk maintain 
the population of 1--5~km objects.

Although larger objects grow throughout this evolution, they accrete a modest 
fraction of the cloud mass. For $q \approx$ 4.0--4.5, the typical 10--20~km 
object doubles its mass every 50--200~Myr. Over the lifetime of Fomalhaut,
continuous capture of material allows the satellites to reach maximum sizes
of roughly 100~km, comparable to the sizes of the irregular satellites of
the giant planets in the solar system \citep[e.g.,][]{bottke2010,kw2011a}.

Within this picture, there are several necessary components for a successful 
capture model with $\ad = \ab \approx 10^{23}$~cm$^2$. 

\begin{itemize}

\item Fomalhaut b must pass through regions of the disk with $d \approx$ 0.3--1.0. 
Otherwise, an Earth-mass planet cannot capture enough material for the nominal
$f_{cap} \approx 2 \times 10^{-3}$. At Fomalhaut b's current position inside the 
orbit of the bright dust belt, the average surface density of the circumstellar 
disk is probably a factor of 3--5 lower than the bright belt \citep{acke2012}. 
Thus, $d \lesssim$ 0.1--0.2.  In this environment, Earth-mass planets may not 
accumulate enough material to produce an observable cross-sectional area of small 
particles.  If Fomalhaut b passes through the dust belt, it encounters regions 
with $d \approx$ 0.5--1.0 and can capture a significant amount of material. Thus, 
the capture model is more viable if Fomalhaut b passes through the dust belt.

\item Captures and collisional evolution within the cloud must maintain a size 
distribution with $q \gtrsim$ 4.0--4.5. Otherwise, planets cannot capture enough
mass to achieve the observed cross-sectional area. Within a standard collisional
cascade, $q \lesssim$ 3.8--3.9 \citep{obrien2003,koba2010a}. However, impacts of
10--100~km objects often produce debris with $q \approx$ 4--6 
\citep[e.g.,][]{durda2004,lein2012}. It seems plausible that a size distribution 
produced from both processes will have an intermediate $q \approx$ 4.0--4.5.

\end{itemize}

Given existing data for Fomalhaut and Fomalhaut b, these conditions are achievable.  
The most likely orbit for Fomalhaut b has $e \gtrsim$ 0.5 and may pass through the 
bright belt of debris \citep{kalas2013,beust2014}. This orbit enables an Earth-mass 
planet to capture material into a large cloud orbiting the planet.  A rough balance
between captures and collisional grinding then yields a cross-sectional area 
$\ad = \ab \approx 10^{23}$~cm$^2$.  Thus, capture is a viable model for dust in Fomalhaut b.

Aside from the ability of an Earth-mass planet to capture sufficient material,
the main uncertainty in this picture is whether captures and collisional grinding
can produce a steep size distribution with $q \approx$ 4.0--4.5.  If these processes
produce a shallower size distribution with $q \lesssim$ 4.0, clouds of captured 
particles will have a much smaller surface area than observed in Fomalhaut b.
We return to these issues in \S4.

\subsection{Collisional Cascade within a Circumplanetary Cloud or Disk}

A collisional cascade is a reliable way to produce a long-lived cloud of dust around a planet
\citep[see, for example,][and references therein]{kw2011a,wyatt2008}.  In this picture, a 
disk or a roughly spherical cloud of solids orbits the planet. Destructive collisions among 
small satellites lead to a cascade of collisions which eventually grinds small particles 
into dust. The largest satellites are often immune to destruction. These satellites may slowly
remove material from the cloud until collisions and radiation pressure remove all of the smaller 
objects.

To explain dust emission in Fomalhaut b, we consider two variants of the collisional cascade
picture. We assume a cloud or disk of material with initial mass $M_d$, particle sizes ranging
from \rmin\ to \rmax, and a power-law size distribution with slope $q$. Following \citep{kw2011a},
collisions drive the evolution. Captures from the circumstellar disk are neglected 
\citep[for an illustration of evolution with an initially massive disk and captures, see][]{bottke2010}.
For either model, a massive swarm of particles ensures a large cross-sectional area for small 
dust grains and a long lifetime for the collisional cascade. Within a circumplanetary disk, a 
large satellite with $R \gtrsim$ 500~km stirs the smaller satellites and maintains high collision
velocities.  Without this satellite, collisional damping among the smaller satellites reduces 
collision velocities and halts the cascade \citep[e.g.,][]{kb2002a}.  Although large satellites
are plausible constituents of a roughly spherical cloud \citep[e.g.,][]{bottke2010,kw2011a}, 
they are not vital for maintaining the cascade.

The properties of a circumplanetary cloud or disk depend on the collision model. The main parameters 
in this model are $q$, \rmax, \rmin, the mass $M_p$ of the planet, the orbital semimajor axis $a$ 
and the eccentricity $e$ of the planet, and the mass \mstar\ and luminosity \lstar\ of the central 
star \citep[e.g.,][and references therein]{kw2011a}.  To explore a large portion of the available 
parameter space with minimal constraints, we consider a simple picture where destructive 
collisions of objects with radii $R$ drive the collisional cascade. We assume all collisions 
produce an array of fragments. In this model, the lifetime of the largest particle is then
the collision time, $t_{coll} \approx \rho R P V / M_d$, where $V$ is the volume of the cloud 
or disk and $P$ is the orbital period. 
Although this approach is much simpler than the collision model in \citet{kw2011a}, it follows
the spirit of more detailed discussions and yields similar results for cloud and disk geometries.

To evaluate this expression, we assign $r$ = 120~AU and \mstar\ = 2~\msun. We assume the swarm
extends to a distance $a_{max} \approx \gamma R_H$ from the planet, where $\gamma \approx$ 0.3 
\citep[e.g.,][]{ham1992,toth1999,shen2008,martin2011}.  Adopting the appropriate volume for a
cloud or disk, we express the collision time in terms of the dust mass:
\begin{equation}
t_{coll} \approx 9 \left ( { R \over {\rm 1~km} } \right ) 
\left ( { {0.01~\mearth} \over M_d } \right ) 
\left ( { \gamma \over 0.3 } \right )^{7/2} 
\left ( { r \over {\rm 120~AU} } \right )^{7/2} 
\left ( { M_p \over {10~\mearth} } \right )^{2/3} 
\left ( { {2~\msun} \over \mstar } \right )^{7/6} ~ {\rm Myr} .
\label{eq: coll-time}
\end{equation}
For swarms containing 1\% of an Earth mass orbiting a 10~\mearth\ planet, destructive collisions 
among objects with $R \approx$ 50--100~km yield lifetimes of 400--800~Myr. Thus, the cascade
can survive for the $\sim$ 400~Myr age of Fomalhaut \citep{mama2012,mama2013}.

Estimating the dust mass in eq. (\ref{eq: coll-time}) requires $q$ and \rmin. Here, we expand
on \citet{kw2011a} and examine several plausible values.  In an equilibrium cascade, the predicted 
slope of the size distribution is $q \approx 3.5-3.7$ \citep[e.g.,][]{obrien2003,koba2010a}. 
To allow some flexibility, we set $q \approx$ 3.5--3.9.  For this range in $q$, most of the 
mass is in the largest objects; most of the cross-sectional area is in the smallest objects. 
As noted in \S3.3, the minimum stable radius for a single particle orbiting a 
1--10~\mearth\ planet is roughly 100--300~\mum\ \citep{kw2011a}.  To examine the impact 
of \rmin, we set \rmin\ $\approx$ 30--300~\mum.

\begin{figure}
\includegraphics[width=6.5in]{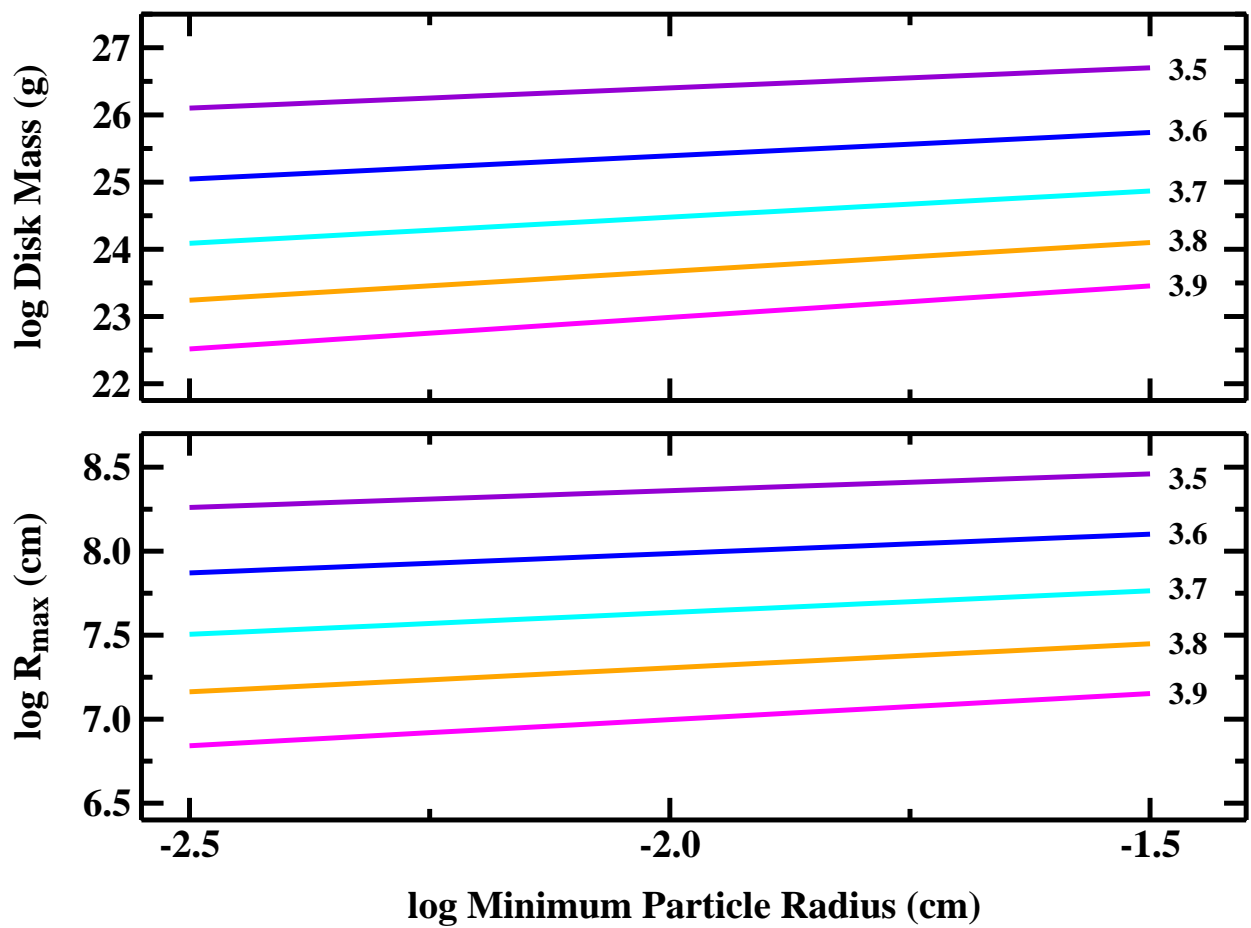}
\vskip 3ex
\caption{%
Circumplanetary disk mass and maximum satellite radius as a function of \rmin\ and 
$q$ for a collisional cascade surrounding a massive planet. Ensembles of particles 
with $q \lesssim$ 3.7 and cross-sectional area $\ad\ = 10^{23}$~cm$^2$ have
the large maximum radius, \rmax\ $\gtrsim$ 300~km, necessary to maintain a
collisional cascade.
\label{fig: casc-mass}
}
\end{figure}

To derive $M_d$, we set $\ad = \ab \approx 10^{23}$~cm$^2$ and calculate \rmax\ as a function 
of $q$ and \rmin\ (eqs. [\ref{eq: area}--\ref{eq: mass}]). Fig. \ref{fig: casc-mass} 
shows the result. For the nominal parameters, $q \approx$ 3.5 and \rmin\ = 300~\mum, the 
mass of the swarm is $M_d \approx$ 0.01~\mearth. The expected collision time is then several 
times the age of Fomalhaut. At fixed $q$, the mass is relatively insensitive to \rmin, 
falling to $M_d \approx$ 0.003~\mearth\ when $\rmin\ \approx$ 30~\mum. At fixed \rmin, 
however, the dust mass is very sensitive to $q$. For reasonable $q \approx$ 3.6 (3.7), 
the mass falls to $M_d \approx$ 0.001~\mearth\ ($M_d \approx 10^{-4}$~\mearth). 
Collision times for swarms orbiting 10~\mearth\ planets are then very long. Because collision
times for small particles are very short, maintaining the cascade is then difficult. However, 
increasing the mass of the central planet shortens the collision time and enables a robust 
collisional cascade throughout the main sequence lifetime of Fomalhaut.

Placing better constraints on the cascade requires a numerical simulation of collisional
evolution in a circumplanetary cloud or disk \citep[e.g.,][]{bottke2010,kb2014}. Although 
evolutionary calculations are cpu intensive, they provide a more robust measure of the size 
distribution, including sizes where $q$ can change dramatically \citep[e.g.,][]{kb2004c}. 
Direct orbit calculations also yield better limits on \rmin\ and the area 
\citep[e.g.,][]{poppe2011}.

Here, we take advantage of the scalability of published calculations to make an independent 
estimate for the collisional lifetime of a circumplanetary disk around Fomalhaut b. As in
eq. (\ref{eq: coll-time}), the lifetime scales with the ratio of the orbital period to the
surface density in the outer disk. Scaling results for circumstellar disks around solar-type
stars \citep[e.g.,][]{kb2008,kb2010} and for circumplanetary disks around Pluto-Charon
\citep{kb2014} yields -- remarkably -- nearly identical time scales (to within a factor of two):
\begin{equation}
t_{coll-scale} \approx 100
\left ( { {0.01~\mearth} \over M_d } \right ) 
\left ( { m \over {10~\mearth} } \right )^{2/3} ~ {\rm Myr} ~ .
\label{eq: coll-time-scale}
\end{equation}
The calculations explicitly derive the growth of large objects; thus, the expression is
independent of $R$. This collision time agrees well with our simple estimate in 
eq. (\ref{eq: coll-time}). Thus, the conclusions derived for the properties of the debris
are robust.

\subsubsection{Summary}

Our simple estimates for the collision time suggest a collisional cascade is a promising
model for dust emission in Fomalhaut b \citep[e.g.,][]{kw2011a,galich2013}. Scaling results 
for the collision time from detailed evolutionary calculations of collisional cascades confirms 
this conclusion. Although continuously replenished during the cascade, the small dust 
particles in a massive circumplanetary debris disk have a large cross-sectional area 
for long time scales.

Our results for \rmax\ and $q$ illustrate the likely range in the mass of a cloud or a disk which 
can produce the measured cross-sectional area in Fomalhaut b. For \rmin\ $\approx$ 100~\mum, the
maximum radius of the size distribution changes from \rmax\ $\sim$ 1000~km for $q$ = 3.5
to \rmax\ $\approx$ 30~km for $q$ = 3.9. The plausible range in the dust mass is equally
large: $\sim$ 0.01~\mearth\ ($q$ = 3.5) to $\sim 10^{-5}$ \mearth\ ($q$ = 3.9). 

These constraints set strong limits on the masses of the central planet 
\citep[e.g.,][]{kw2011a,galich2013}. For $q$ = 3.5, swarms with 0.01~\mearth\ of solid 
material orbiting a 10~\mearth\ planet produce the observed area in Fomalhaut b for
the likely main sequence lifetime of Fomalhaut. Although increasing the slope of the 
size distribution to $q = $ 3.7 (3.9) enables smaller masses, long collision lifetimes
require a more massive planet, $m \approx 100~\mearth$ ($m \approx 1000~\mearth$). 
Current near-IR observations allow sub-Jupiter mass planets, but not super-Jupiter mass
planets \citep{janson2012,currie2012,currie2013}. Thus, current data preclude systems
with $q \gtrsim 3.9$.

\section{DISCUSSION} 

In \S3, we considered three generic models -- impacts, captures, and collisional
cascades -- for the origin of a cloud of dust in Fomalhaut b. In the simplest model,
a single giant impact within the circumstellar disk produces an expanding cloud of 
dust orbiting the central star. As another simple alternative, dynamical processes 
during the earliest stages of planet formation leave a massive cloud or disk of solid
particles surrounding a planet. Collisions among the largest satellites maintain a swarm 
of dust particles around the planet. The capture model is an interesting combination of
these ideas, where a planet continuously captures the debris from giant impacts 
within its Hill sphere. If the cloud of debris becomes massive enough, a 
balance between material gained through capture and lost by a collisional cascade 
sets the properties of the circumplanetary dust cloud. Each of these models makes
predictions for the mass and cross-sectional area of the dust cloud. Our analysis 
in \S3 establishes these predictions.

\begin{figure}
\includegraphics[width=6.5in]{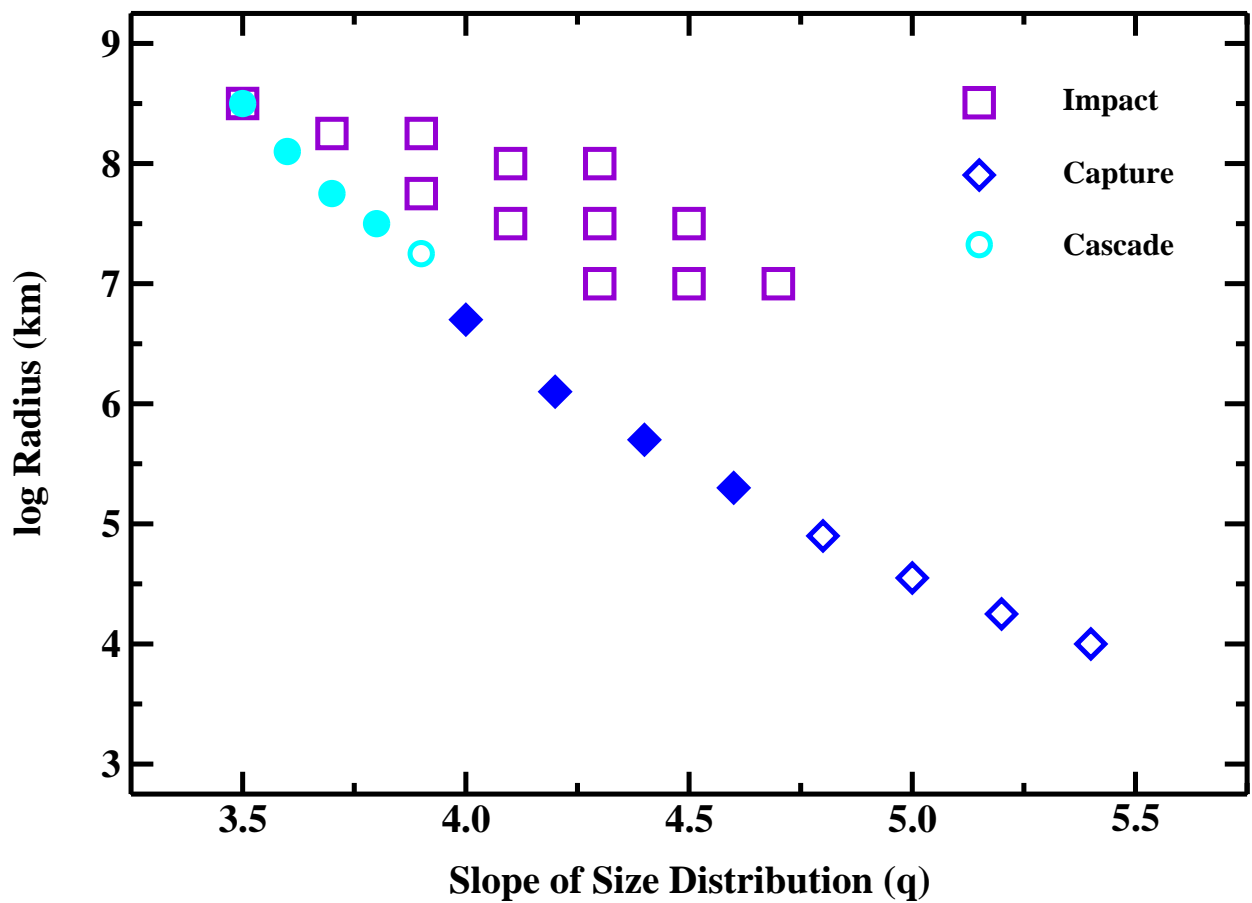}
\vskip 3ex
\caption{%
Combinations of $q$ and radius (\rmax\ for captures and cascades; 
$R_1$ for impacts) which can produce the observed cross-sectional area 
$\ab \approx 10^{23}$~cm$^2$ for models of giant impacts (violet points), 
captures (blue points), and collisional cascades (cyan points).  Open 
symbols show combinations of $q$ and $\rmax, R_1$ which match the observed 
area.  Filled circles indicate physically plausible combinations. Among
allowed impact models, collisions are either too rare (large $R_1$) or
require unlikely $q$ (small $R_1$). For captures and cascades, estimates
for the cloud lifetime favor small $q$ and large \rmax. 
\label{fig: summary}
}
\end{figure}

To summarize the constraints on each model, we collect the derived parameters for 
the slope of the dust size distribution ($q$) and the maximum radius of the size 
distribution (\rmax, for captures and cascades) or the radii of two impactors ($R_1$). For 
simplicity, we consider steps of 0.2 in $q$ and 0.25 in log radius. The open 
symbols in Fig.~\ref{fig: summary} show combinations of $q$ and $\rmax, R_1$ 
which match the observed $\ab \approx 10^{23}$~cm$^2$. 

Although the allowed parameter space is broad, simple physical arguments limit
the parameter space considerably. For giant impact models (\S3.2), collisions 
among pairs of objects with $R \gtrsim$ 100~km happen too rarely. Collisions 
among smaller objects occur more often, but standard collision outcomes produce
debris with too little cross-sectional area to match observations. Non-standard 
outcomes with little debris in large particles can match the observed area with
large $q$. Current numerical experiments of collisions suggest this option is
improbable \citep[e.g.,][]{durda2004,lein2012}. Thus, giant impacts seem an 
implausible way to produce a dust cloud in Fomalhaut b.

Capture models appear somewhat more viable (\S3.3). Earth-mass planets orbiting 
Fomalhaut at 120~AU can attract up to $10^{23}$~g of solids in 100~Myr.  If this 
material maintains a steep size distribution, then the cross-sectional area of 
the cloud matches observations of Fomalhaut b. Although many combinations of
$q$ and \rmax\ yield a model \ad\ which can match the observed \ab\ for 
\md\ $\approx 10^{21} - 10^{23}$~g,
the collision time precludes models with $q \gtrsim$ 4.6. When $q$ is too large,
the largest particles have short collision times. Short collision times limit 
the mass of the cloud to $M_d \lesssim 10^{21}$~g, which is insufficient to 
produce the observed \ab\ with \rmin\ $\approx$ 10--1000~\mum.  With this 
constraint, we limit the allowed parameter space to the four filled diamonds
in Fig.~\ref{fig: summary}.

Collisional cascade models are also reasonable \citep[\S3.4, see][]{kw2011a}. 
Within the allowed parameter space, size distributions with \rmax\ $\gtrsim$ 500~km
can maintain the cascade for the age of Fomalhaut. In systems with smaller \rmax\ 
and larger $q$, there is too little material in the most massive objects. Thus, the
cascade cannot survive for the 200--400~Myr age of Fomalhaut.  Discounting these 
options limits the allowed parameter space to the three filled circles in 
Fig.~\ref{fig: summary}. 

Within $(q, \rmax)$ space, there are two main regions. Collisional 
cascade models permit $q \approx$ 3.5--3.7 and $\rmax \approx$ 500--3000~km.  Systems 
with smaller $q$ require larger \rmax.  Capture models allow $q \approx$ 4.0--4.6 and 
$\rmax \approx$ 2--50~km. Systems with smaller $q$ require larger total mass.  Both of 
these pictures require 1--10 Earth-mass planets.  Our analysis strongly favors these 
options over a giant impact. Plausible giant impacts occur too rarely, require 
unlikely collision outcomes, or both.

These conclusions generally agree with previously published results. For giant impacts,
\citet{kalas2005} and \citet{tamayo2013} derive similarly low probabilities for 
collisions among 100--1000~km objects.  Although \citet{galich2013} revise the collision 
probability upward, their estimate is based on the surface density of material within 
the belt. Given the newly measured trajectory of Fomalhaut b \citep{kalas2013,beust2014} 
and the short lifetime of the debris cloud, any giant impact capable of producing 
Fomalhaut b must occur at distances $r \lesssim$ 120~AU where the surface density is 
at least a factor of six smaller than in the belt (\S2.1).  Thus, the \citet{galich2013} 
estimate of the collision frequency is overly optimistic. 

Compared to \citet{galich2013}, our approach to the outcomes of high velocity collisions 
between two protoplanets yields more ejected mass but less surface area.  By using 
approximations appropriate for cratering collisions between a small object and a much 
larger one, \citet{galich2013} underestimate dust production from collisions between 
objects with roughly equal masses (\S3.2).  With the ejected mass known, \citet{galich2013} 
set \rmin, \rmax, and $q = 3.5$ to yield the observed area.  Our estimates for \ad\ hinge 
on numerical experiments which derive the size of the largest fragment as a function of 
the ejected mass. After associating the size of the largest fragment with \rmax, we derive
\ad\ as a function of $q$.  Despite the larger ejected mass, this approach yields much 
larger \rmax\ and much smaller \ad.  Given current collision theory 
\citep[e.g.,][and references therein]{lein2012}, our results seem more realistic.  
Discriminating between the two methods requires new numerical experiments of high velocity 
collisions.

Coupled with recent dynamical results, our collision analysis in \S3.2 enables stronger
limits on the impact hypothesis. \citet{tamayo2013} infers that collisions between 
two large objects are unlikely to lead to the large $e$ orbit in Fomalhaut b.  He favors 
a collision between a small planetesimal and a much larger protoplanet already on a large 
$e$ orbit. However, collisions between one small and one large object produce enough dust 
only when the large object has $R \gtrsim$ 1000~km (\S3.2). These collisions are very 
unlikely.  Along with the need to produce the apparent apsidal alignment of Fomalhaut b 
and the main belt, these constraints challenge our ability to develop a viable impact 
model \citep[e.g.,][]{tamayo2013,beust2014}. 

Capture models applied to Fomalhaut b have a limited history.  \citet{kw2011a} consider 
capture of material which strikes the central planet and ejects dust from the planet's 
surface. Based on our analysis, we agree with their conclusion that the cross-section of
a 1--10~\mearth\ planet is too small to accrete enough mass for the Fomalhaut b dust cloud.
Our results in \S3.3 generally confirm their estimates for the amount of mass ejected in
the collision. In our picture, the larger cross-section of the Hill sphere enables a larger 
capture rate.  Both approaches ignore likely captures from circumstellar material striking 
orbiting satellites \citep{durda2000,stern2009,poppe2011}; this process likely adds captured 
material to the circumplanetary environment. Addressing the viability of this model in more 
detail requires numerical simulations.

Finally, we agree with previous studies supporting the collisional cascade model
\citep{kw2011a,galich2013,kalas2013,tamayo2013}. Most studies derive similar properties 
for the central planet, 1--100~\mearth, and the surrounding circumplanetary cloud, $\sim$ 
0.01~\mearth. The stability, surface area, and lifetime of the cloud set the lower
mass limit on the planet \citep{kw2011a,galich2013}; minimizing disruption of the main 
dust belt sets the upper mass limit \citep{chiang2009,kw2011a,tamayo2013,beust2014}. 
Our approach expands the allowed range of slopes for the size distribution of particles
in a circumplanetary cloud or disk. Because the slope correlates with the dust mass, 
future dynamical studies can provide additional constraints on these parameters. 

To explore the available parameter space for these models in more detail, we now 
examine plausible uncertainties (\S4.2), tests (\S4.3), and improvements (\S4.4) 
of our approach. 

\subsection{Uncertainties}

\subsubsection{Observations}

To examine how uncertainties impact our results, we begin with the derivation of 
the cross-sectional area from the observations. As outlined in \S2, we assume that
all radiation from Fomalhaut b is scattered light from Fomalhaut. The minimum 
cross-sectional area is then derived from the ratio of the scattered flux to the
flux from Fomalhaut. The uncertainty is these quantities is small, $\sim$ 10\%.
Thus, the uncertainty in the minimum cross-sectional area is small. 

Establishing an upper limit on the cross-sectional area requires an accurate estimate
for the optical depth. Our analysis in \S2.5 safely precludes $\tau \gtrsim$ 1 for 
giant impact models. Optically thick clouds orbiting a massive planet have collision 
times roughly $10^{10}$ times shorter than the age of Fomalhaut, robustly eliminating 
this possibility. With $\tau \lesssim$ 1, the observed \ab\ yields an accurate estimate 
of the true \ab.

Deriving the true cross-sectional area of the dust requires an estimate of the 
albedo $Q$. Among Kuiper belt objects in the solar system, the albedo is typically 
$Q \approx$ 0.04--0.20 \citep{marcialis1992,roush1996,stansberry2008,brucker2009}.
Choosing $Q \approx$ 0.1 thus yields a reasonable estimate for the actual
cross-sectional area, $\ab \approx 1.25 \times 10^{23}$~cm$^2$, with a factor of 
two uncertainty.

This uncertainty has little impact on our results 
(e.g., Fig.~\ref{fig: area-mdust}). For configurations with large \rmax/\rmin,
changing \ab\ by a factor of 2 modifies $M_d$ by a factor of $2^{2/3}$ = 1.6.
For giant impacts with fixed $q$, this uncertainty implies a 20\% variation in 
the derived target radius, a factor of two difference in the collision rate,
and minimal revision to our conclusions. If the target radius is held fixed,
a factor of two uncertainty in \ab\ implies a 0.1--0.2 change in $q$.  We infer 
similar adjustments to $q$ and \rmax\ for captures or collisional cascades. Thus,
allowing for observational error in the cross-sectional area leads to minimal
changes in the allowed parameter space of Fig.~\ref{fig: summary}. 

\subsubsection{Size Distribution}

On the theoretical side, we assume that the size distribution is a power law with 
a slope $q$ and a clear minimum size \rmin\ and maximum size \rmax. Adopting a 
single largest remnant in a giant impact is reasonable.  In a collisional cascade, 
the largest objects resist erosion by accreting smaller objects 
\citep[e.g.,][]{kb2008,kb2010,kb2012}.  For the slopes inferred from our analysis, 
the next two largest objects have radii $R \approx$ 0.75--0.85~\rmax\ and 
$R \approx$ 0.65--0.75~\rmax.  Thus, a single largest object is appropriate for
impact, capture, and cascade models.

Establishing the proper \rmin\ is somewhat more involved.  When giant impacts 
yield small dust grains orbiting Fomalhaut at $r \approx$ 100~AU, setting the 
minimum radius equal to or larger than the blowout radius -- \rmin\ $\gtrsim$ 
5~\mum\ -- is sensible. If small ($R \lesssim \rmin$), icy grains at 120~AU 
have impurities of carbon or silicates, radiation pressure probably ejects 
them on the orbital or a smaller time scale \citep{arty1988,gust1994}.  


Independent of their total mass, grains with $R \lesssim \rmin$ probably contain 
a large fraction of the cross-sectional area of the ejecta. With velocities much 
larger than the escape velocity of the impactors, they produce a rapidly expanding 
halo around the main ejecta. While visible for several years, very small grains 
become invisible on time scales much longer than a decade \citep[e.g.,][]
{galich2013,kalas2013}.

For capture and cascade models, isolated small particles are ejected when
radiation pressure overcomes the gravity of the planet. For $M_p \approx$ 1--10~\mearth,
\rmin\ $\approx$ 300~\mum\ \citep{burns1979,kw2011a}. Although smaller particles
might participate in the collisional processing of either mechanism, typical
collision times are much longer than the planet's orbital period.  Thus, particles 
with $R \lesssim \rmin$ leave after several orbits of the planet around Fomalhaut
\citep{poppe2011}.  

For impact models, adding more complexity to the size distribution is not warranted. 
As long as there is a broad range of sizes between \rmin\ and \rmax, a single 
power-law provides a reasonably good way to relate the cross-sectional area, the 
mass, and the parameters -- $q$, \rmin, and \rmax\ -- of the size distribution. 
Thus, this uncertainty seems minimal.

For viable capture models, a single power-law may not completely characterize the 
size distribution from 100~\mum\ to 50--100~km. In our picture, capturing the 
fragments of giant impacts yields a steep size distribution with $q \gtrsim 4$ 
\citep[e.g.,][]{durda2004,lein2012}. Collisional evolution among fragments tends
to produce shallower size distributions with $q \approx$ 3.5--3.7 
\citep{obrien2003,koba2010a}.  While a single power-law may not capture all details 
of capture and collisional evolution, it is probably sufficient to establish 
allowed values for $q$ and \rmax.

In collisional cascades, the proper equilibrium size distribution is uncertain
\citep[e.g.,][and references therein]{bely2011}. However, it is somewhat 
inaccurate to adopt a single power-law to describe the numbers of objects from 
a few microns to a few thousand kilometers. In long-term numerical simulations 
of cascades, the size distribution is better represented by separate power-laws 
at small ($ R \lesssim$ 0.1--1 ~km), intermediate ($R \approx$ 1--100~km), and large 
sizes \citep[$R \gtrsim$ 10--100~km; e.g.,][]{kb2004c,kbod2008,bottke2010,kb2012}. 
Analytic studies support this conclusion \citep{pan2005,schlicht2013}. Wavy
patterns are often superimposed on these power-laws \citep[e.g.,][]{campo1994,
obrien2003,bely2011}.  The slopes of the power laws for the small and large objects 
are similar, with $q_S \approx 3.5-4.0$ and $q_L \approx$ 2.5--4.5; the slope of 
the intermediate power law is small, with $q_I \approx$ -1 to 1.  Observations of 
Kuiper belt objects in the solar system reveal fairly strong evidence for a break 
in the size distribution at $\sim$ 20--100~km \citep[e.g.,][]{fuentes2008,fraser2010b} 
and some evidence for another break at small radii \citep[e.g.,][]{schlicht2012}. 
Observed slopes are generally consistent with theoretical predictions 
\citep{bottke2010,kw2011b,schlicht2012}.

Quantifying how a somewhat wavy, multi-component power law approximation to the 
size distribution impacts our conclusions requires exploring a vast parameter space. 
To place an initial limit, we examine a few 
general cases for a typical outcome, $q_S \approx$ 3.5--4.0 and $q_I \approx$ 1. 
Compared to models of a single size distribution with $q \approx$ 3.5--3.7 and
\rmax\ $\approx$ 500--3000~km, multiple power laws with $q_L > q_S$ match the 
observed \ab\ with an \rmax\ which is larger by a factor of 1.5--2.  When 
$q_L < q_S$, matching \ab\ requires a steeper slope, with 
$q_S \approx q + \delta q$ and $\delta q \approx$ 0.4--0.5. Because the 
intermediate part of the size distribution contains little area or mass, 
multiple power law models require a factor of 2--4 more mass to achieve the
same surface area.

Although a factor of 2--4 uncertainty in the mass certainly impacts the size
distribution and the lifetime of a collision cascade 
\citep[e.g.,][]{wyatt2008,kb2008,kriv2008,kw2010,kb2010}, it has little impact 
on the general viability of collisional cascade models. 
From eq.~(\ref{eq: coll-time}), the 
collision time scales inversely with the mass and linearly with $M_p^{2/3}$. 
For a fixed cascade lifetime, changing the mass involved in the cascade requires 
a corresponding adjustment to $M_p$.  With current observations requiring 
$M_p \lesssim 0.5 M_J$, it is fairly straightforward to adjust the mass required
for the collisional cascade and meet the broad range of allowed planet masses.

\subsubsection{Collision Physics}

Aside from the physical parameters of the size distribution, the physics of collisions 
and collision outcomes plays a major role in our analysis. Our results for impacts 
hinge on understanding debris production during collisions between large objects. 
Analyses of captures and cascades also rely on swarms of solid particles finding 
stable equilibrium size distributions over long periods of time.

Although each mechanism depends on an accurate parameterization of the binding energy
of icy objects, the uncertainties in \qdstar\ probably have little impact on our results.
In a collisional cascade, satellites lose mass when the collision energy exceeds 
\qdstar\ \citep[e.g.,][]{dohn1969,obrien2003,kb2008,koba2010a}.  Because $Q_c$ varies 
with the orbital velocity, it is possible to compensate for changes in \qdstar\ simply 
by changing the mass of the planet. For a reasonably large range in planet masses, 
the subsequent evolution of the cascade is largely unchanged.

The binding energy has little impact on capture models. In our picture, \qdstar\ helps
to set the collision time for the largest objects (eqs. [\ref{eq: rd1}--\ref{eq: t-large}]).
Although factor of two uncertainties in \qdstar\ can lead to similar uncertainties in the
rate particles lose mass, the collision time for the largest objects depends mainly on the
cross-sectional area. Thus, our assumptions for \qdstar\ have a relatively small, 
$\sim$ 10\% to 20\%, impact on the collision time and the total mass of the cloud.

Viable impact models are more sensitive to \qdstar.  Changing \qdstar\ by a factor of 
two changes debris production by a similar factor. Less debris (larger \qdstar) makes
giant impacts less viable. Although more debris adds to the viability of giant impacts,
these models still require steep size distributions with little mass in the largest
remnant. These outcomes are still unlikely.

Our assumption of a head-on giant impact has little impact on our conclusions. 
When impacts are off-center, the center-of-mass impact energy is smaller by a factor
$b$, the impact parameter \citep[e.g.,][]{asphaug2006,lein2012,sstew2012}. 
With less energy available in a collision and the same energy required to unbind half 
the colliding protoplanets, off-center collisions lose less mass than head-on collisions.
Thus, allowing for off-center collisions reduces the likelihood that a giant impact is
responsible for the dust in Fomalhaut b.

For captures and cascades, the outcomes of collisions have little impact on the results.
For sizes where collisions produce destruction or growth, the rate particles diminish
or grow depends on the collision rate much more than collision outcomes 
\citep{koba2010a,koba2010b,koba2011}. Simple physics constrains the collision rates.

Our conclusions for giant impacts rely heavily on the published outcomes of numerical 
experiments of high energy collisions \citep[e.g.,][]{benz1999,durda2004,lein2012}. 
So far, different approaches yield similar results: high velocity collisions between
objects with substantial self-gravity always leave behind large remnants with a
significant fraction of the debris. Because the dust cloud in Fomalhaut b appears to
require debris with little mass in the largest remnants, a large impact seems an unlikely
way to produce the dust cloud.
If numerical simulations identify collision parameters capable of producing the debris
required in Fomalhaut b, an impact becomes much more plausible.

\subsubsection{Orbital Dynamics}

Finally, several aspects of orbital dynamics might modify our conclusions. 
In a planet with a highly elliptical orbit, for example, the size of the 
Hill sphere is smaller at periastron than at apoastron. Satellites with 
circumplanetary orbits at semimajor axes of 0.3--0.4~\rhill\ might be 
stable at apoastron but unstable at 
periastron. Because these satellites have orbital periods comparable to 
the orbital period of Fomalhaut b around Fomalhaut, it should take many 
Fomalhaut b orbits to develop unstable satellite orbits 
\citep[e.g.,][]{shen2008}. Indeed, \citet{kalas2013} showed that planets 
with $M_p \gtrsim 5 \times 10^{24}$ g on Fomalhaut b-like orbits can retain 
satellites with semimajor axes of 0.3--0.4 \rhill.  Thus, the elliptical 
orbit has little impact on the stability of captured satellites or satellites 
involved in a collisional cascade.

Dynamical interactions among satellites also play a role in the viability
of capture and collisional cascade models. In an ensemble of satellites,
gravitational interactions produce random velocities comparable to the
escape velocity of the largest satellite \citep[e.g.,][]{gold2004}. When
these random velocities exceed the orbital velocity, satellites are ejected.
In the transneptunian region of the solar system, these interactions 
produce the scattered disk -- an ensemble of 10--500~km icy objects with
perihelia near the orbit of Neptune and large orbital eccentricity
\citep[e.g.,][]{gladman2008}.

Managing the cascade around 1--10~\mearth\ planets with much larger 
satellites is challenging.  Massive satellites with $R \approx$ 
500--1500~\kms\ have escape velocities, $v_{esc} \approx$ 0.5--1.5 \kms, 
much larger than the local orbital velocity. On a few dynamical time scales,
these objects eject smaller satellites orbiting within 2--3 Hill radii 
\citep[e.g.,][]{glad1993,gold2004}, which is roughly 0.05--0.06 AU. For a 
satellite system with an outer radius of 0.25--0.5~AU, 5--10 massive 
satellites can eject all small objects on very short time scales. 

Maintaining a roughly spherical cloud of dust in a capture or cascade model 
is also challenging. For any initial geometry, energy loss and angular 
momentum transport from inelastic collisions eventually produce a prograde 
disk with angular momentum similar to the initial angular momentum of the 
cloud  \citep{brah1976}. If the initial orbits within the cloud are roughly
balanced between prograde and retrograde, material gradually falls onto the
planet instead of landing in a large disk. This evolution probably enhances 
the mass loss rate from a roughly spherical collisional cascade, shortening
the lifetime.

\subsection{Tests}

The simplest ways to deduce the source of the optical emission in Fomalhaut b
involve polarimetry or spectroscopy. Imaging polarimetry excels at probing the 
underlying geometry of dusty clouds or disks \citep[e.g.,][]{whit1993,whit1997,olof2012}.
Optical or IR spectroscopy might reveal absorption features from the central
A-type star \citep{lagr1995,hemp2003} or silicate features from dust 
\citep{teles1991,wein2003}.  Measuring the velocity of cloud material with high 
resolution spectroscopy \citep[e.g.,][]{olof2001,brand2004} would discriminate
between expanding and orbiting geometries.  Unfortunately, these observations are 
far in the future.  The HST and the {\it James Webb Space Telescope} (JWST) 
have no polarimetric capabilities. 
Although the source is too faint for HST spectroscopy, the prototype exposure 
time calculator\footnote{ http://jwstetc.stsci.edu/etc/input/nirspec/spectroscopic/} 
for NIRSPEC on JWST yields an 8$\sigma$ detection for an A-type continuum with 
an exposure time of 3600 sec.  Although JWST is scheduled for launch no sooner 
than 2018, low resolution NIRSPEC spectra may enable accurate tests of dust models 
for Fomalhaut b.

Extending photometry to longer wavelengths also tests these models 
\citep[e.g.,][]{currie2012,currie2013}. JWST NIRCAM observations will enable better 
than 10$\sigma$ detections\footnote{http://jwstetc.stsci.edu/etc/input/nircam/imaging/}
at 1--3~\mum\ \citep[e.g.,][]{tamayo2013}.
On a somewhat longer time scale, ground-based imaging with 20-m to 40-m class telescopes
might provide independent measures of the spectral energy distribution at 1--5~\mum.  

Until JWST launches, other approaches are possible. To develop tests for giant impact 
models, we assume an unbound cloud of dust particles with an expansion velocity exceeding 
the escape velocity of a pair of impactors with $R \approx$ 50~km, 
$v_{esc} \approx 5 - 6 \times 10^3 ~ \cms$. 

\begin{itemize}

\item Expansion of the cloud is detectable on short time scales. If all of the 
material expands at $v_{esc}$, the expansion rate is roughly 0.01~AU~yr$^{-1}$ 
\citep[e.g.,][]{galich2013,kalas2013}. However, the ejecta probably have a
range of velocities with $f(>v) \propto (v / v_{esc})^{-\alpha}$ 
\citep{gault1963,stoff1975, housen2003,housen2011}. Adopting $\alpha$ = 1.5, 
roughly 20\% (50\%) of the material has $v \gtrsim 3 ~ v_{esc}$ ($1.6 ~ v_{esc}$).
If 35\% of the dust expands at twice $v_{esc}$, the diameter grows roughly 
0.4~AU (1--2 pixels on HST images) in 10~yr.  Although current efforts to 
resolve the source are inconclusive \citep[e.g., \S2;][]{galich2013,kalas2013}, 
improving the resolution and placing robust limits on the expansion rate should 
be possible in the next decade \citep[e.g.,][]{tamayo2013}.

\item Shearing of the cloud is also detectable. For particles expanding at 
median velocity $v$ from a guiding center with orbital velocity $v_K$, the
velocity dispersion is roughly $\delta v \approx v$ 
\citep[e.g.,][]{gault1963,housen2003}. Among particles expanding tangentially
to the orbital motion, some lag the orbit; others move ahead of the orbit. 
Thus, the sphere shears into a ring \citep[e.g.,][]{kb2005}.  When
$\delta v / v_K \approx v / v_K \approx 0.01 - 0.02 $, the differential motion 
is $\delta r / r \approx 0.01 - 0.02$.  Over 10 yr, the guiding center moves 
roughly 8~AU \citep{kalas2013}, resulting in a predicted shear of 0.15--0.3~AU.  
In the next decade, HST and JWST data can test this prediction.

Our large estimate for the shearing rate -- a few decades instead of 100--1000~yr 
\citep{currie2012,galich2013,kalas2013,tamayo2013} -- is based on the larger internal 
velocity dispersion of debris clouds suggested by laboratory and numerical experiments. 
Performing SPH simulations of collisions between pairs of 50--200~km objects 
\citep[e.g.,][]{lein2012} in a Keplerian reference frame would test these ideas.

\item Collisions with other circumstellar disk particles enhance these rates. 
For a typical relative velocity of 0.4~\kms, it takes roughly 10~yr for a disk
particle to cross the cloud.  During this period, one in $10^3$ disk particles 
collides with a cloud particle.  If the surface density of the disk at 
$r \approx$~120~AU is 1\% to 10\% ($d = 0.01 - 0.10$) of the initial surface density, 
roughly $10^{21}$~g to $10^{22}$~g of disk material mixes with particles in the 
expanding cloud every decade.  Because orbits in the disk differ from orbits of 
the cloud, these collisions enhance the rate of expansion and orbital shear by 
factors of three or more.

Even very modest amounts of mixing -- $\sim 10^{18}$~g, corresponding to a disk 
with $d \approx 10^{-5}$ -- can easily increase the expansion and shear by 
roughly 50\%.  Any mixing thus increases the chances of detecting expansion or 
shearing very soon. If Fomalhaut b enters the dust belt \citep{galich2013,kalas2013},
the enhanced collision rate should produce an obvious shear on very short time scales.

\end{itemize}

Without high quality polarimetry or spectroscopy, robust tests of the capture or 
cascade pictures are more challenging.  Still, several tests allow promising tests
of either scenario.

\begin{itemize}

\item Although significant expansion or contraction of captured or cascading material
is unlikely, measurements of Fomalhaut b's size place important constraints on the 
models. As noted in \S2, unambiguous resolution of the disk of Fomalhaut b places a
robust lower limit on the mass of a central planet. 

\item Placing better limits on the azimuthal structure of material at 20--130~AU 
also constrains models for Fomalhaut b. If dust in the inner disk is smoothly 
distributed \citep[as in the model of][]{acke2012}, capture models are more viable.
Detecting patchy dust increases the likelihood of massive planets in the inner disk
and decreases the likelihood of significant capture of small solids by planets in
the inner disk.

\item Collisional cascades should leave behind a trail of small particles 
\citep[e.g.,][]{kalas2013}. Because
particles with $R \approx$ 5--300~\mum\ are blown out of circumplanetary -- but not
circumstellar -- orbits, these particles should take up orbits along the path of
Fomalhaut b. Assuming 10$\sigma$ detections from existing observations of Fomalhaut b, the 
brightest detectable cloud of small dust particles is a factor of 5--10 fainter than 
Fomalhaut b.  With roughly 1000 resolution elements along the elliptical orbit, it is possible 
to discriminate a trail from the background if an ensemble of small dust particles has a 
total cross-sectional area $A_{d,s} \approx 1000 / 5-10 \approx$ 100--200 times the cross-sectional 
area of the dust observed in Fomalhaut b.  If it is possible to coadd data convincingly 
in an annulus along the orbit, a robust algorithm could detect fainter trails.

Although there are many uncertainties, this level of emission from 5--300~\mum\ particles
is plausible.  For size distributions with $q \approx$ 3.5--4.0, the cross-sectional area 
of the small particles is 7--20 times larger than the area of the circumplanetary debris disk. 
If the particles do not drift too far away from the orbit and if collisions do not destroy 
small particles ejected well before the current epoch, it is possible to enhance 
this surface area by factors of 3--10 
\citep[see, for example][and references therein]{wyatt2008,kb2008,kb2010,kw2010,kw2011a}. 
Unambiguous limits on this trail would enable stern tests of capture and cascade models.

\item Fomalhaut b's possible entry into Fomalhaut's dust belt provides another opportunity 
to test cascade \citep[e.g.,][]{kalas2013} and capture models. By analogy with Saturn's
rings \citep[e.g.,][]{dur1989} and Kuiper belt objects \citep[e.g.,][]{stern2009}, we
expect several classes of behavior when particles from the dust belt interact with
circumplanetary dust: (i) large objects from the belt will carry away small circumplanetary
particles and (ii) collisions between small belt objects and large circumplanetary objects
will produce debris. For small objects in a captured cloud or disk, entry into the dust
ring will be dramatic: we expect an initial loss of captured material on 10 yr time scales, 
followed by a slow increase as the rare collisions of larger objects produce debris which
repopulates the smaller sizes.  Because collisional cascades have a shallower dust distribution,
we expect much less dramatic changes: as large objects remove small particles from the cloud
or the disk, collisions between larger objects rapidly restore lost material. The time scale
for any variations, however, should be similar, $\sim$ 10--100~yr.

\end{itemize}

\subsection{Improvements}

As observations continue to probe the nature of Fomalhaut b, new approaches can hone 
theoretical predictions. Although clear improvements in analytic approaches are possible,
here we outline several numerical calculations to clarify expectations.

For all dust models, it is crucial to add to our understanding of interactions between 
the dust cloud and ambient material in the disk. If the surface density of the disk at
30--100~AU is roughly 1\% to 10\% of the initial surface density, then disk material inevitably 
interacts with the cloud. From our earlier estimates, disk material with a total mass 
comparable to the mass of the cloud interacts with cloud material every 10--100 yr (for 
impact and capture models) to every $10^4 - 10^5$ yr (for cascade models). If these
interactions add material to the cloud, they make (i) capture and cascade models more 
viable and (ii) impact models less viable. Interactions which remove material from
the cloud tend to decrease the viability of all models. 

Despite the wealth of analytic and numerical work 
\citep[e.g.,][]{dones1993,kort2005,jewitt2007a,nesv2007,
pires2012}, estimating the amount of material a planet can capture throughout the history
of a planetary system remains uncertain. For Fomalhaut b, its elliptical orbit through the
inner disk and main belt of dust might lead to substantial differences in the capture rate.
Numerical simulations can place stronger constraints on our simple estimates.

Numerical experiments could clarify the long-term collisional evolution of 
circumplanetary debris. Our estimates for the viability of the capture hypothesis
rest on the development of an equilibrium between the rate captures add mass to the
cloud and the rate collisions remove mass from the cloud.  More sophisticated
calculations can address this issue. 

Finally, numerical calculations can illuminate the relative importance of cloud and 
disk geometries for collisional cascades around massive planets. During the early 
evolution of the solar system, dynamical interactions between the gas giants strongly 
favor cloud geometries for swarms of captured particles \citep[e.g.,][]{nesv2007}. In 
a dynamically quiet environment, however, growing planets might easily capture large 
disks of particles. Unless these disks are disrupted by the gravity of another massive 
planet, massive satellite formation and the onset of a collisional cascade is inevitable.
Understanding common features and differences of circumplanetary cascades in clouds and
disks might enable new tests of the cascade model.

\section{SUMMARY}

We explore the ability of three generic models -- giant impacts, captures, and 
collisional cascades -- to account for a large dust cloud in Fomalhaut b. After
deriving the basic observational constrains (\S2), we develop a novel approach to
the power-law size distribution of solid particles (\S3.1) and apply this technique
to giant impacts (\S3.2), captures (\S3.3), and collisional cascades (\S3.4). Despite
the uncertainties in our approach (\S4.1), we derive several clear constraints.

\begin{enumerate}

\item Giant impacts seem the least plausible model 
\citep[for another approach for rejecting this picture, see][]{tamayo2013}. 
Although it is possible to produce enough debris in a collision between two 
100~km objects, the outcome required to match the observations is unlikely. 
However, this model is the easiest to test: simple theory expects detectable 
expansion and shearing of the belt in the next decade 
\citep[e.g.,][]{galich2013,kalas2013}.

\item Although capture models are viable, it is challenging to produce a stable cloud or 
disk of captured particles which lasts for the main sequence lifetime of Fomalhaut.  
If small particles contain most of the mass, planets can capture enough material from 
the inner disk and main belt of Fomalhaut to produce the observed cross-sectional area.  
We speculate that these systems can find an equilibrium size distribution from captures
and collisional evolution.  More detailed calculations can test this idea. 

\item A collisional cascade is the least problematic model. In principle, Earth-mass or 
larger planets with reasonable circumplanetary reservoirs of solid material can maintain 
a cascade for the main sequence lifetime of Fomalhaut. However, it is unclear whether 
the system can reach the required equilibrium between damping and stirring 
\citep[e.g.,][]{kw2011a}.  

\item Although testing the capture and cascade models, both should leave detectable trails 
of small particles along their orbits. Current data -- or images acquired with JWST -- might 
reveal these trails.

\end{enumerate}

Within the next decade, observations with HST or JWST can test these models. Detecting 
image expansion/shear or a trail of small particles along Fomalhaut b's orbit is possible
with either facility.  IR spectroscopy with JWST would provide a direct measure of the 
amount of scattered light from a dust cloud. If Fomalhaut b passes through the main belt,
comparing time variations in brightness with the predictions of more detailed theoretical
calculations should also constrain the models.

Whatever the nature of Fomalhaut b, it gives us new insights into collisional evolution
and planet formation theory.  As the observational constraints grow, we are likely to 
learn more about the outcomes of individual collisions and the evolution of swarms of
solid particles.

\vskip 6ex

Comments from M. Geller and a thoughtful referee improved our presentation.
Portions of this project were supported by the {\it NASA } {\it Astrophysics Theory}
and {\it Origins of Solar Systems} programs through grant NNX10AF35G and the {\it NASA}
{\it Outer Planets Program} through grant NNX11AM37G.

\bibliography{ms.bbl}

\clearpage

\begin{deluxetable}{llcccc}
\tablecolumns{6}
\tablecaption{Fomalhaut b Data}
\tabletypesize{\footnotesize}

\tablehead{{Telescope/Instrument}&{Filter}&{$\lambda$ ($\mu m$)}&{Apparent Magnitude}&{Flux Density ($\mu$Jy)}
& {Reference}}
\startdata
HST/ACS&F435W & 0.435 & 25.22 $\pm$ 0.18 & 0.32 $\pm$ 0.06 & 1\\
"& F606W & 0.606 & 24.95 $\pm$ 0.13 & 0.36 $\pm$ 0.04& 1\\
"& F814W & 0.814 & 24.91 $\pm$ 0.20 & 0.27 $\pm$ 0.05& 1\\
HST/STIS & 50CORO& 0.574 & 24.96 $\pm$ 0.20 & 0.38 $\pm$ 0.08 & 2\\ 
HST/WFC3&F110W &1.15&-- & $<$ 1.60 & 3\\
Subaru/IRCS&J & 1.25 & $>$ 22.22 & $<$ 3.36 & 1\\
Keck/NIRC2&H & 1.65 & $>$ 22.60 & $<$ 0.94 & 4\\
Spitzer/IRAC&$[3.6]$ & 3.6 & $>$ 13.82&$<$ 833& 5\\
Spitzer/IRAC&$[4.5]$ & 4.5 & $>$ 16.66&$<$ 38.80 & 6\\
Spitzer/IRAC&$[5.8]$ & 5.8 & $>$ 11.32&$<$ 3416& 5\\
Spitzer/IRAC&$[8.0]$ & 8 & $>$ 11.12 &$<$ 2317& 5\\
 \enddata
\tablecomments{References: 1) \citet{currie2012}, 2) This work, 3) \citet{galich2013}, 
 4) \citet{currie2013}, 5) \citet{maren2009}, 6) \citet{janson2012}.  
Upper limits for non-detections are quoted as 5-$\sigma$ upper limits.  
For the F606W photometery, we use the average of the 2004 and 2006 measurements: 24.92 $\pm$ 0.10 
and 24.97 $\pm$ 0.09. The \textit{HST/STIS} filter central wavelength refers to the pivot wavelength listed 
in the \textit{fits} header.}
\label{fombphot}
\end{deluxetable}

\begin{deluxetable}{lccc}
\tablecolumns{4}
\tablecaption{Fomalhaut b Spatial Extent Estimates in ACS Data}
\tabletypesize{\footnotesize}
\tablehead{{Data Set} & {~~FWHM (pt. source)~~} & {~~FWHM (linear fit) [x,y]~~} & {~~FWHM (gaussian fit) [x,y]~~} \\ 
{} & {milliarcsec} & {milliarcsec} & {milliarcsec}
}
\startdata
F435W & 50 & [56 $\pm$ 15, 69 $\pm$ 14] & [59 $\pm$ 13, 71 $\pm$ 18] \\ 
F606W (2004) & 69 & [75 $\pm$ 13, 80 $\pm$ 17] & [68 $\pm$ 13, 75 $\pm$ 13] \\
F606W (2006) & 69 & [69 $\pm$ 14, 57 $\pm$ 15] & [69 $\pm$ 14, 56 $\pm$ 15] \\
F814W & 94 & [120 $\pm$ 60, 65 $\pm$ 25] & [108 $\pm$ 25, 60 $\pm$ 30] \\ 
\enddata
\label{fwhmest}
\end{deluxetable}
%

\clearpage

\clearpage

\clearpage

\clearpage

\clearpage

\clearpage

\clearpage

\clearpage

\clearpage

\clearpage

\clearpage

\clearpage

\end{document}